\documentclass[aps,twocolumn,pra,10pt,superscriptaddress,showpacs,floatfix]{revtex4-2}
\usepackage{amsmath}
\usepackage{dsfont}
\usepackage{amssymb}
\usepackage{physics,esint}
\usepackage{graphicx}
\usepackage{appendix}
\usepackage{color}
\usepackage{url}
\usepackage[dvipsnames]{xcolor}
\usepackage[colorlinks=true,linkcolor=Blue,urlcolor=Blue,citecolor=Blue]{hyperref}

\begin{document}
\title{Ultimate tradeoff relation of quantum precision limits in multiparameter linear measurement}
	
\author{Guolong Li}
\email{glli@hdu.edu.cn}
\author{Xiao-Ming Lu}
\email{lxm@hdu.edu.cn; http://xmlu.me}
\affiliation{Department of Physics, Hangzhou Dianzi University, Hangzhou 310018, China}
\affiliation{Zhejiang Key Laboratory of Quantum State Control and Optical Field Manipulation, Hangzhou Dianzi University, Hangzhou 310018, China}
\date{\today}
	
\begin{abstract}
Linear measurements are widely applied in sensing classical signals, e.g., gravitational wave (GW), and are developing toward joint measurement of multiple parameters.
In this work, focusing on multiparameter linear measurements of classical monochromatic signals, we establish an inherent tradeoff relation that tightly constrains the quantum limits on estimation precision.
The tradeoff relation is fundamental since it is rooted in Heisenberg's uncertainty principle, and fully characterizes the dependence between the attainable precision limits on the estimated parameters.
Eventually, we identify a necessary condition under which an optimal measurement protocol saturates the tradeoff relation, and show that the measurement phase can be regulated to implement flexible allocation of precision weights.
Our finding can offer valuable guidance for detuned GW sensors in ultra-sensitive searches for post-merger remnants.
\end{abstract}
	
\maketitle
	
\section{Introduction}
In quantum sensing and metrology, \textit{linear measurement}, defined as a linear-response sensing model distinct from standard POVM measurement in quantum information, has been widely applied in detecting, e.g., gravitational waves (GW) \cite{Braginsky1992,Danilishin12, RevModPhys.86.121}, dark matter \cite{PhysRevLett.132.101001}, infrared rays \cite{PhysRevLett.132.153602}, and rotation velocity~\cite{Li22JOSAB39}.
To study quantum noise, linear-response theory has been developed for linear devices \cite{Kubo1966}, such as quantum-limited force/displacement sensors \cite{Braginsky1992}.
Moreover, apart from constructing waveform estimation \cite{Tsang2011} and distributed quantum sensing \cite{Luca2025}, multiparameter estimation has incorporated linear measurement to gain more information. 
Especially, detuned LIGO-like interferometers is a profound example.
Compared with $\sim{100}$~Hz GWs, the kilohertz ones are favorable to reveal the post-merger remnant of binary neutron-star mergers, and thus can reflect the otherwise inaccessible cores of neutron stars to better constrain the neutron-star equation-of-state and enable other explorations \cite{BAIOTTI2019103714, PhysRevLett.122.061102, universe7040097, PhysRevD.79.044030, PhysRevX.4.041004, Ott2009, Lasky_2015}.
While standard interferometric GW detectors have become vital linear sensors after the great success of LIGO \cite{PhysRevLett.116.061102, AdvancedLIGO, AdvancedVirgo, KAGRA, LIGOVirgoKAGRA, PhysRevX.9.031040, PhysRevX.11.021053, science.abc7397, Miller2019, 10.1093/nsr/nwx029}, detuned LIGO-like interferometers have attracted much attention due to their outperformance of detecting kilohertz signals.
However, based on Heisenberg's uncertainty principle (HUP), the detuning frequency in these linear detectors leads to incompatibility of the optimal measurements for two independent parameters of a monochromatic signal
\footnote{We only thoroughly discuss monochromatic signals because, as long as the dominant monochromatic components are precisely detected, the waveform of signals can be uncovered. 
In any frequency $\Omega$, only independent parameters $A$ and $B$ in the monochromatic signal $s(t) = A \cos \Omega t + B \sin \Omega t$ should be estimated and reveal the information in the waveform.} \cite{gardner2023holevo, PhysRevLett.132.130801}.
Thus, quantum multiparameter estimation theory is essential for achieving quantum-limited linear measurement \cite{HELSTROM1967101, Helstrom1054108, Yuen1055103, Belavkin1976, HelstromBook1976, HolevoBook1982, Personick1054643, HayashiBook2005, PhysRevX.10.031023}.

In classical parameter estimation, the Cram\'er-Rao bound (CRB) characterizes attainable accuracy, quantified by the Fisher information matrix (FIM) \cite{Fisher1922, Fisher1925, KayBook1993, WassermanBook2010, CasellaBook2002, LehmannBook1998}.
In contrast to single-parameter estimation \cite{BRAUNSTEIN1996135, Fujiwara2006}, simultaneous multiparameter estimation cannot guarantee asymptotic attainability of quantum CRB owing to measurement incompatibility from HUP in quantum regime \cite{BUSCH2007155, Belliardo2021}.
As a compromise, many prior error bounds are formulated in terms of weighted mean estimation errors \cite{Carollo2019, Rubio2018,  tsang2021holevo, albarelli2020upper, Sidhu2020, PhysRevA.101.022303, Sidhu2021, PhysRevA.94.052108, PhysRevLett.116.180402, PhysRevLett.120.030404, Suzuki2016, Suzuki2019, Suzuki2020, Kull2020, CAROLLO20201, PhysRevA.61.042312, Nagaoka2005, Matsumoto2002}.
Among them, the Holevo Cram\'er-Rao bound (HCRB) has been expected to establish the limit in multiparameter linear measurement \cite{PhysRevLett.123.200503, PhysRevLett.132.130801}.

In this work, we employ the information regret tradeoff relation (IRTR) \cite{PhysRevLett.126.120503}, and establish an ultimate tradeoff relation of the precision limits on incompatible parameters in multiparameter linear measurements of classical monochromatic signals (Here “ultimate” is restricted to the precision bound within the linear measurement framework, rather than a universal bound for arbitrary measurement schemes).
Compared with HCRB adopted in previous work \cite{PhysRevLett.132.130801}, the proposed tradeoff bound sheds further insights into the incompatible measurement.
As a tight bound in a tradeoff form, our result truly identifies the boundary of attainable errors in an analytical form.
On the other hand, HCRB still implies a tight bound since the convex envelope of the HCRB family coincides with our proposed bound, as elaborated in the following discussion.
However, thanks to its explicit analytical form, our adopted tradeoff bound bypasses the cumbersome optimization and convex-envelope construction required for HCRB, and further offers an intuitive picture of the inherent multiparameter precision tradeoff curve/surface.
Furthermore, the IRTR, as the foundation of our bound, can be saturated for arbitrary pure states, while the HCRB is saturable when parameters act as the mean shift of a pure Gaussian state \cite{HolevoBook1982}.
The HCRB is still asymptotically attainable via many-copy quantum estimation due to quantum local asymptotic normality, yet collective measurements are generally needed \cite{PhysRevA.94.052108}.
By contrast, The saturation of IRTR is attainable with only separable measurements.
Therefore, our framework can be readily generalized to measurement setups beyond the linear model and yield saturable bounds with separable measurements.

We further find that, even if there exists incompatibility, a measurement scheme is capable of saturating the IRTR under an optimal condition.
Moreover, we demonstrate that, in this measurement scheme, both attainable estimation errors of two parameters strictly accord with the saturated IRTR and are adaptable via a probe phase.
Finally, we intuitively show that, while detuned frequency in laser GW interferometers enhances the sensitivity of kilohertz GW signals, it leads to visible tradeoff effect.
It reveals that, as the detuned device is considered for precisely detecting kilohertz signals, one has to think over the tradeoff relation to investigate possible improvements of the quantum measurements.

\section{Basic framework}
\begin{figure}[tbhp]
\includegraphics[width=0.7\columnwidth]{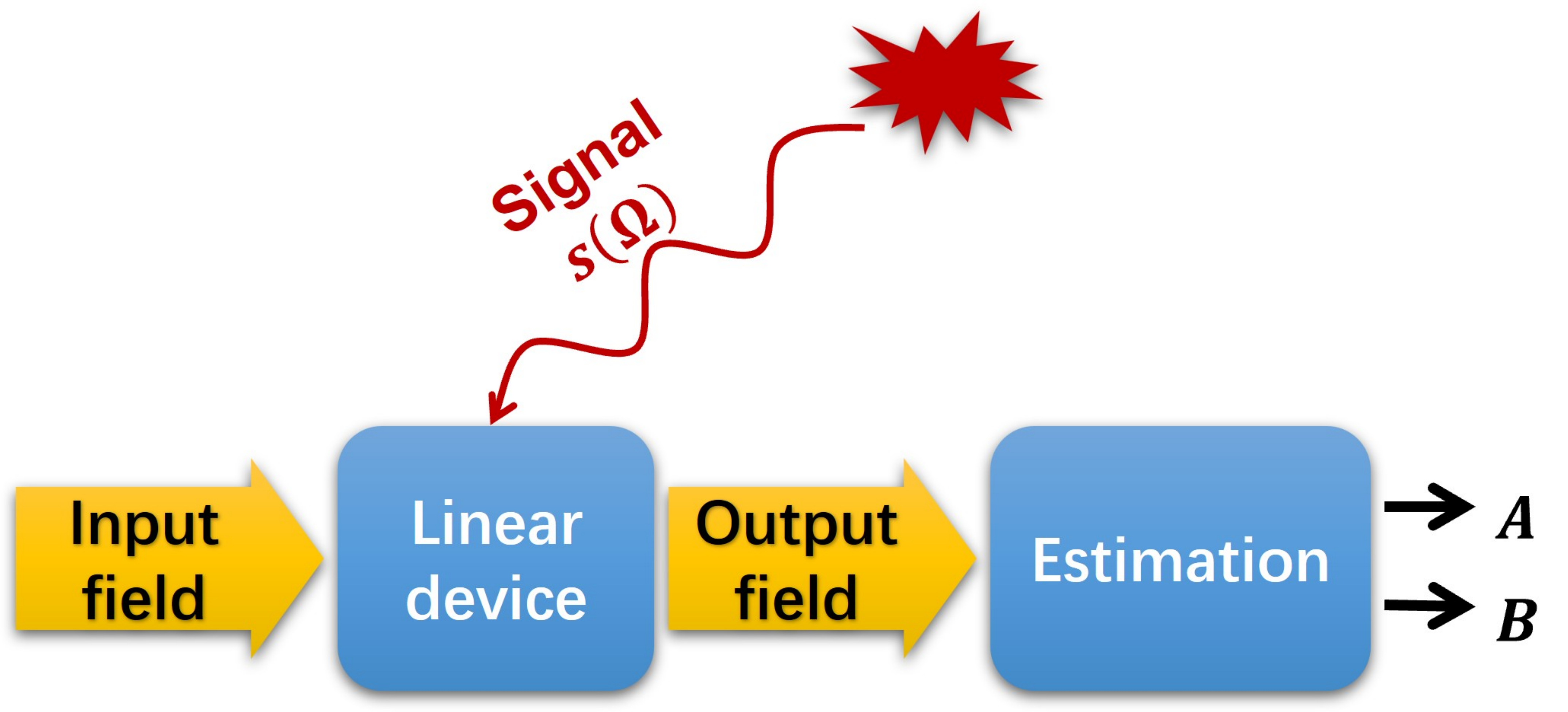}
\centering
\caption{
    Schematic representation of the multiparameter linear measurement for sensing a monochromatic signal $s(\Omega) = T (A + iB) /2$ with parameters of interest $A$ and $B$.
}
\label{fig:schematics}
\end{figure}

We first introduce a general linear device.
As presented in Fig.~\ref{fig:schematics}, input port observable ${G}$ of the linear device, as a {\it generator}, is linearly coupled to the waveform signal $s(t)$ via the Hamiltonian $H_\mathrm{int} = s(t) {G}$.
Consequently, any observables that do not commute with the generator ${G}$ linearly respond to signal $s(t)$ \cite{PhysRevLett.119.050801, PhysRevA.95.012103} and, based on the input/output relation \cite{PhysRevA.31.3761}, directly transfer the signal information to detectable output modes usually denoted by dimensionless canonical quadratures $x$ and $p$ with $[{x},{p}] = i$.
In the Fourier domain, the detected output field $z=x, p$ is given by 
${z}(\Omega) = {z}^{(0)} (\Omega) + \chi_{z G} (\Omega) s(\Omega)$   \cite{Kubo1966, PhysRevD.65.042001, PhysRevLett.132.130801},
where ${z}^{(0)}$ is the corresponding free quadrature without signal impacting and $\chi_{z G} (\Omega)$ is the susceptibility determined by the commutator of any operator ${z}$ and the generator ${G}$.
For a monochromatic dimensionless signal $s(t) = A \cos (\Omega t) + B \sin (\Omega t)$ at a positive angular frequency $\Omega$,
the linear measurement is explicitly expressed as \cite{PhysRevLett.132.130801}
\begin{equation}\label{eq:linear system}
    \left[
    \begin{array}{c}
	{x}(\Omega) \\
	{p}(\Omega)
    \end{array}
    \right] =
    \left[
    \begin{array}{c}
	{x}^{(0)}(\Omega) \\
	{p}^{(0)}(\Omega)
    \end{array}
    \right] +
    \left[
    \begin{array}{c}
	\chi_{xG} (\Omega) \\
	\chi_{pG} (\Omega)
    \end{array}
    \right]
    \frac{T}{2} (A + i B),
\end{equation}
with Fourier transformation $s(\Omega) = T (A + i B) /2$ of monochromatic signals $s(t)$ in finite integration time $T$.

The real and imaginary parts of $x(\Omega)$ and $p(\Omega)$ define a Hermitian operator vector as
$\mathbf {x}=(x_1, p_1, x_2, p_2) := \sqrt{\frac{2}{T} } (\Re[x(\Omega)], \Re[p(\Omega)], \Im[x(\Omega)], \Im[p(\Omega)])$.
The elements of $\mathbf{x}$ encode parameters $A$ and $B$ as
\begin{equation}\label{eq:vec x}
    {\mathbf x}=
     {{\mathbf x}}^{(0)} + A \mathbf{d}_A + B \mathbf{d}_B,
\end{equation}
and obey the canonical commutation relation as 
$[x_j ,p_k] = i \delta_{jk}$ and $[x_j ,x_k] = [p_j ,p_k] = 0$, as do the elements of  the initial quadrature vector ${\mathbf x}^{(0)} := \qty(x_1^{(0)}, p_1^{(0)}, x_2^{(0)}, p_2^{(0)})  $ without signal impacting.
Here  
$\mathbf{d}_A = \sqrt{\frac{T}{2}} (\Re [\chi_{xG}], \Re [\chi_{pG}], \Im [\chi_{xG}], \Im [\chi_{pG}])^\mathrm{T}$
and 
$\mathbf{d}_B = \sqrt{\frac{T}{2}} ( -\Im [\chi_{xG}], -\Im [\chi_{pG}], \Re [\chi_{xG}], \Re [\chi_{pG}])^\mathrm{T}$
are two orthogonal vectors with a common Euclidean norm 
$\mathcal{N} = \sqrt{ \frac{T}{2} \left(\left| \chi_{xG} \right|^2 + \left| \chi_{pG} \right|^2\right)}$. 
Obviously, the parameters of interest $A$ and $B$ displace two independent harmonic oscillators with canonical quadrature $(x_j, p_j)$ ($j=1,2$).
If the initial $\mathbf{x}^{(0)}$ is set as a vacuum state, corresponding to the covariance matrix of $\mathbf{x}^{(0)}$ being $\mathrm{diag} (\frac{1}{2}, \frac{1}{2}, \frac{1}{2}, \frac{1}{2})$, the parameters $A$ and $B$ are thus yields a two-mode coherent state 
$ \ket{\beta_1, \beta_2} = D (\beta_1) \ket{0} \otimes D (\beta_2) \ket{0}$ 
with displacement operator
$D(\beta)$ and the amplitudes
\begin{align}\label{eq:beta}
\beta_1 &= \frac{1}{\sqrt{2}} 
\qty[
(d_{A,1} A + d_{B,1} B)  
+ i (d_{A,2} A + d_{B,2} B) 
], \notag \\
\beta_2 &= \frac{1}{\sqrt{2}}
\qty[
(d_{A,3} A + d_{B,3} B)  
+ i (d_{A,4} A + d_{B,4} B)
],
\end{align}
where $d_{A,j} \ (d_B,j)$ is the $j$th element of the vector $\mathbf{d}_{A} \ (\mathbf{d}_B)$.
Moreover, the na{\"i}ve optimal unbiased estimates
$\hat{A}_\mathrm{nv} = (\mathbf{d}_A \cdot \mathbf{x})/\mathcal{N}^2$ and 
$\hat{B}_\mathrm{nv} =( \mathbf{d}_B \cdot \mathbf{x}) /\mathcal{N}^2 $ of $A$ and $B$ satisfy 
\begin{equation}\label{eq:naive commu}
  \qty[ \hat{A}_\mathrm{nv},  \hat{B}_\mathrm{nv}] = \frac{i \mu}{\mathcal{N}^2}, 
\end{equation}
where 
$\mu := T \mathcal{N}^{-2} 
(
\Re\left[ \chi_{pG}  \right] 
\Im\left[ \chi_{xG}  \right]   
- \Re\left[ \chi_{xG} \right] \Im \left[ \chi_{pG}  \right]
)$ ($0\leq \mu \leq 1$) 
has been introduced in Ref.~\cite{PhysRevLett.132.130801}.

In quantum multiparameter estimation theory, the density operator ${\rho}$ that depends on an unknown vector parameter ${\vec{\theta}} = (\theta_1, \theta_2, ..., \theta_n)$ can be exploited to estimate $\vec{\theta}$ via positive-operator-valued-measure (POVM)
${M} = \{ {M}_\omega | {M}_\omega \geq 0, \ \sum_\omega {M}_\omega = \mathds{1} \}$.
Then an outcome $\omega$ is obtained with probability $p(\omega) = \mathrm{tr} ({M}_\omega {\rho})$ based on Born's rule in quantum mechanics.
If $\hat{\theta}_j$ is an estimator for parameter $\theta_j$ to map observation data to the estimates, the error-covariance matrix $\mathcal{E}$ is defined as
$\mathcal{E}_{jk} := \mathbb{E}_\theta[(\hat{\theta}_j-\theta_j) (\hat{\theta}_k-\theta_k)] $, 
where the expectation $\mathbb{E}_\theta [\bullet]$ is taken with respect to the probability of observation data.
In unbiased estimation, it satisfies the CRB as
$\mathcal{E} \geq F(M)^{-1}$ \cite{Cramer1946, Rao1992}
\footnote{Here the relation of two matrices $X \geq Y$ means positive semi-definite matrix $X - Y$.}
with the classical FIM
\begin{equation}\label{eq:CFIM}
	F({M})_{jk} = \mathbb{E}_\theta \left[\frac{\partial \ln p(\omega)}{\partial \theta_j} \frac{\partial \ln p(\omega)}{\partial \theta_k} \right].
\end{equation}
Obviously, classical FIM depends on POVM ${M}$ via probability $p(\omega)$ and, thus, characterizes the performance of a quantum multiparameter measurement.

Quantum parameter estimation pursues the optimization of estimation precision over quantum measurements.
The quantum CRB manifests that, for any quantum measurement ${M}$, the classical FIM is bounded as \cite{PhysRevLett.72.3439, Liu_2020}
\begin{equation}\label{eq:QCRB}
	F({M}) \leq \mathcal{F} \implies \mathcal{E} \geq  F^{-1} \geq  \mathcal{F}^{-1},
\end{equation}
with quantum FIM $\mathcal{F} = \Re \mathcal{Q}$.
Here the Hermitian matrix $\mathcal{Q}$, called {\it quantum geometric tensor}, is defined as
$\mathcal{Q}_{jk} := \mathrm{tr} ({L}_j {L}_k {\rho})$ via Hermitian symmetric logarithmic derivative (SLD) ${L}_j$ which obeys
$({L}_j {\rho} + {\rho} {L}_j)/2 = \partial {\rho} / \partial \theta_j$. 
The latter inference is obvious when classical CRB is considered.
It implies that, for multiple estimated parameters, a joint measurement attains optimality only if its classical FIM saturates quantum FIM.
However, if the optimal measurements for individual parameters are incompatible, this saturation is impossible, that is, no joint measurement can simultaneously achieve the individual quantum CRB of each parameter.
To reflect this case in quantum multiparameter estimation, due to Eq.~\eqref{eq:QCRB}, a positive semi-definite matrix is defined as
$R({M}) := \mathcal{F} - F({M})$,
called regret of Fisher information.
Subsequently, the normalized-square-root regret is defined as 
$\Delta_j := \sqrt{R_{jj} / \mathcal{F}_{jj}} 
= \sqrt{ ( \mathcal{F}_{jj} - {F}_{jj} ) / \mathcal{F}_{jj} }$
in an interval $[0, 1]$.
For incompatible measurement, the information regrets of corresponding parameters cannot simultaneously vanish, but satisfy the IRTR as \cite{PhysRevLett.126.120503}
\begin{equation}\label{eq:IRTR}
	\Delta_j^2 + \Delta_k^2 + 2\sqrt{1-c_{jk}^2} \Delta_j \Delta_k 
	\geq c_{jk}^2,
\end{equation}
with \textit{incompatibility coefficient}
\begin{equation}\label{eq:incompat coeff}
	c_{jk} = \frac{| \Im \mathcal{Q}_{jk} |}{\sqrt{\Re \mathcal{Q}_{jj} \Re \mathcal{Q}_{kk}}}
	= \frac{| \Im \mathcal{Q}_{jk} |}{\sqrt{\mathcal{F}_j \mathcal{F}_k}}.
\end{equation}
If the state ${\rho}$ is pure, the inequality \eqref{eq:IRTR} is tight and there exist measurements that turn it into an equality.
Notably, even though the first inequality in Eq.~\eqref{eq:QCRB} is untight for incompatible measurement, Eq.~\eqref{eq:IRTR} still allows us to acquire a tight ultimate tradeoff between precisions of estimated parameters.
The general framework, owing to its unique advantages \cite{Yung2024}, is favorable to analyze the joint linear measurement for two signal parameters.

\section{Quantum precision limit}
Based on the pure output state $\ket{\beta_1, \beta_2}$ with amplitudes in Eq.~\eqref{eq:beta}, the quantum geometric tensor $\mathcal{Q}$ with respect to the parameters $A$ and $B$ is given by (see Appendix \ref{appd:mat Q} for details)
\begin{equation}\label{eq:mat Q}
	\mathcal{Q}=2 \mathcal{N}^2 \left(
	\begin{array}{cc}
		1       & i\mu \\
		-i\mu & 1
	\end{array}
	\right).
\end{equation}
Thus, the quantum FIM and the corresponding incompatibility coefficient, defined by Eq.~\eqref{eq:incompat coeff}, should be
\begin{equation}\label{eq:mat qF}
	\mathcal{F}=\Re \mathcal{Q} =
        2 \mathcal{N}^2 \left(
	\begin{array}{cc}
		1   & 0 \\
		0   & 1
	\end{array}
	\right), \quad
     c_{12} = \mu,   
\end{equation}
respectively.
Based on the CRB Eq.~\eqref{eq:QCRB} and our result Eq.~\eqref{eq:mat qF}, the individual quantum CRB from single-parameter estimation is $\mathcal{E}_{A (B)} \geq 1/(2 \mathcal{N}^2)$ where $\mathcal{E}_{A(B)}$ denotes the variances of the parameters $A(B)$.
However, owing to HUP which leads to incompatible quantum measurements, the variances of any simultaneous measurement can not attain the individual quantum CRB, but should obey IRTR Eq.~\eqref{eq:IRTR}.

Combined with our results Eqs.~\eqref{eq:mat qF}, the IRTR Eq.~\eqref{eq:IRTR} indicates that there exist a family of optimal measurements extracting the Fisher information such that the regret inequality
$\Delta_1^2 + \Delta_2^2 +2\sqrt{1-\mu^2} \Delta_1 \Delta_2 \geq \mu^2 $
is tight.
Accordingly, the IRTR explicitly reads
\begin{equation}\label{eq:CFIM tradeoff relation}
	\begin{split}
		&\left(2 - \frac{F_{11} + F_{22}}{2 \mathcal{N}^2}\right) \\
		&\ +2 \sqrt{1-\mu^2} \sqrt{
			\left(1 - \frac{F_{11}}{2\mathcal{N}^2} \right)
			\left(1 - \frac{F_{22}}{2\mathcal{N}^2} \right)
		}
		\geq \mu^2.
	\end{split}
\end{equation}
Owing to CRB Eq.~\eqref{eq:QCRB}, it is transformed into
\begin{equation}\label{eq:error tradeoff relation}
	\begin{split}
		&\qty[2 - \frac{ (\mathcal{N}^2 \mathcal{E}_{A}) ^{-1}+ (\mathcal{N}^2 \mathcal{E}_{B}) ^{-1}}{2}]\\
		&\ +2 \sqrt{1-\mu^2} \sqrt{
			\qty[1 - \frac{ (\mathcal{N}^2 \mathcal{E}_{A})^{-1}}{2} ]
			\qty[1 - \frac{ (\mathcal{N}^2 \mathcal{E}_{B})^{-1}}{2} ]
		}
		\geq \mu^2.
	\end{split}
\end{equation}
This inequality is one of our main results since it fully reflects the tradeoff between parameter estimation variances $\mathcal{E}_A$ and $\mathcal{E}_B$, especially for the optimal measurements that make it tight.

The incompatibility coefficient $c_{12}$, defined in Eq.~\eqref{eq:incompat coeff}, plays a key role on the tradeoff relation and is exactly equal to $\mu$ from Eq.~\eqref{eq:naive commu}, i.e., $c_{12} = \mu$ as Eq.~\eqref{eq:mat qF}. 
For $\mu = 0$, both precisions can simultaneously attain the individual quantum limits
$\mathcal{E}_{A(B)} = 1/(2 \mathcal{N}^2$).
However, as the incompatibility coefficient $\mu$ rises, Eq.~\eqref{eq:error tradeoff relation} indicates that both variances $\mathcal{E}_A$ and $\mathcal{E}_B$ are dependent mutually and thus can not simultaneously attain their individual quantum CRBs.

On the other hand, another widely studied CRB, called Holevo Cram\'er-Rao bound (HCRB), also provides a lower bound on variances in quantum multiparameter  estimation \cite{Xia2023}. 
The previous work \cite{PhysRevLett.132.130801} that proposed the linear sensing model has adopted  HCRB as
	$\Sigma_H =  \min_{\phi \in (0,\pi]} 
	\frac{1}{\mathcal{N}^2} \left(\frac{w}{\cos (\phi)^2}
	+
	\frac{1-w}{\cos [\phi + \arcsin (\mu)]^2}
	\right)$
from the jointly coherent state, where $w \in (0, 1)$ represents the weight to construct \textit{weighted mean squared estimation error}
$ \Sigma =  2 w \mathcal{E}_A + 2 (1-w) \mathcal{E}_B $.
This framework implies that the estimation errors $\mathcal{E}_{A(B)}$ fulfills 
\begin{equation}\label{eq:HCRB relation}
	2 [w \mathcal{E}_A  + (1-w) \mathcal{E}_B]
	\geq \Sigma_H.
\end{equation}
So far, there are two distinct proposals for describing the bounds of the estimation errors, i.e., the IRTR Eq.~\eqref{eq:error tradeoff relation} and the HCRB Eq.~\eqref{eq:HCRB relation}.

\begin{figure}[tbhp]
	\includegraphics[width=1\columnwidth]{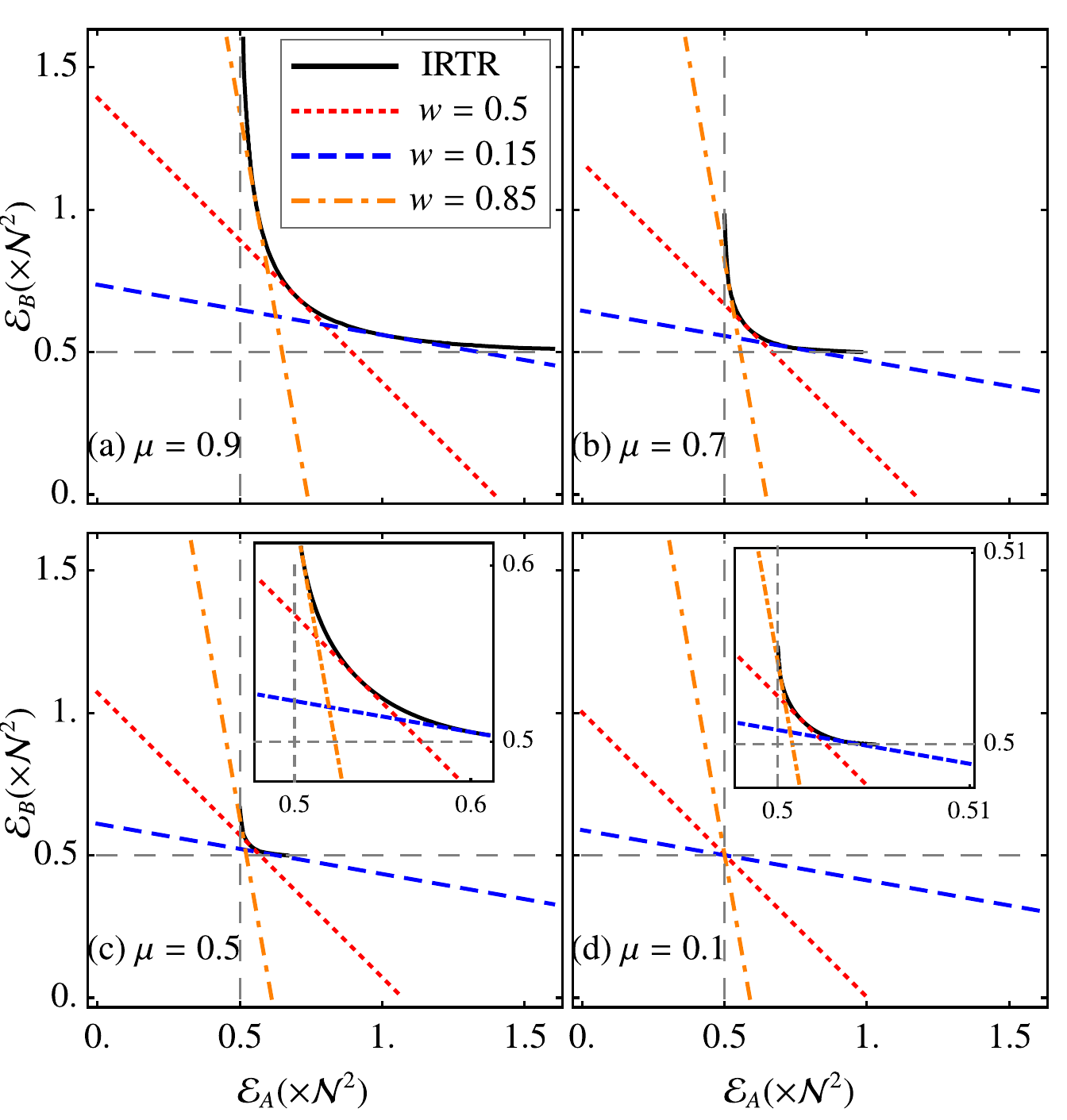}
	\centering
	\caption{
        Comparison between the IRTR and the HCRB with respect to parameters $A$ and $B$ in monochromatic signals, for (a) $\mu=0.9$, (b) $\mu = 0.7$, (c) $\mu=0.5$, and (d) $\mu = 0.1$.
		In all panels, the lines include the IRTR (black solid line) and the HCRBs with $w = 0.5$ (red dotted line), $w=0.15$ (blue dashed line), and $w=0.85$ (orange dot-dashed line).
	Each inset of panels (c) and (d) shows the corresponding zoom-in.
	}
	\label{fig:Comp 4 Miu}
\end{figure}

As plotted in Fig.~\ref{fig:Comp 4 Miu}, the comparison between IRTR and HCRB for multiparameter linear sensing reveals that IRTR, employed in our work via Eq.~\eqref{eq:error tradeoff relation}, is more informative than HCRB with specific weight.
Clearly, HCRBs with various weights $w$ are exactly tangent to the IRTR, implying that the convex envelope of HCRB family is equivalent to the IRTR.
However, the IRTR directly captures the exact interdependence between two estimation errors and full information regarding the attainable error region, while the HCRB additionally requires tedious optimization and convex-envelope construction.
Furthermore, Fig.~\ref{fig:Comp 4 Miu} shows that the line of IRTR with larger $\mu$ is significantly more curved.
Hence, our result in Eq.~\eqref{eq:error tradeoff relation} explicitly represents how $\mu$ as an incompatibility coefficient reinforces the tradeoff relation.

\section{Measurement protocol}
After an appropriate symplectic transformation $\mathcal{M}$, the quadrature vector ${\mathbf x}$ in Eq.~\eqref{eq:vec x} is transformed into the other one as
${\mathbf X} = \mathcal{M} \mathbf{x} := ({X}_1, {P}_1, {X}_2, {P}_2 )^\mathrm{T} $ that remains the same  canonical commutation relation (see Eq.~\eqref{eq:vec X} in Appendix \ref{appd:sys}).
If the initial state is a vacuum state, it is equivalent to a two-mode coherent state $\ket{\alpha_1, \alpha_2}$ with amplitudes $\alpha_1$ and $\alpha_2$ in the Schr{\"o}dinger picture (see Eq.~\eqref{eq:alphas} in Appendix \ref{appd:sys}).
The adopted compatible unbiased estimates are given by 
\begin{align}\label{eq:measurement}
	\hat{A} &= \mathcal{N}^{-1} 
    [{X}_1 - \mathcal{T}(\phi) {P}_2],
	\notag \\ 
	\hat{B} &= \mathcal{N}^{-1}
    [- \mathcal{S}(\phi) {P}_1 +  \mathcal{C}(\phi) {X}_2],
\end{align}
for jointly estimating both parameters $A$ and $B$, where the coefficients are
\begin{align}
	\mathcal{C} (\phi) &= \cos \phi / \left( \sqrt{1-\mu^2} \cos \phi - \mu \sin \phi \right), \notag \\
	\mathcal{S} (\phi) &= \sin \phi / \left(\sqrt{1-\mu^2} \cos \phi - \mu \sin \phi\right),
\end{align} 
and $\mathcal{T} (\phi) = \mathcal{S} (\phi) / \mathcal{C} (\phi) =  \tan \phi$, relying on a phase $\phi$.
Obviously, $[\hat{A}, \hat{B}] = 0$, and both observable expectations are $\expval{\hat{A}}_0 = A$ and $\expval{\hat{B}}_0 = B$ when $\expval{\bullet}_0 = \expval{\bullet}{0}$ takes the vacuum state $\ket{0}$.

Based on Eq.~\eqref{eq:CFIM} with the simultaneous eigenstate of measurement operators Eq.~\eqref{eq:measurement}, we can obtain the classical FIM as
\begin{equation}\label{eq:mat cF}
	F=2 \mathcal{N}^2 \left[
	\begin{array}{cc}
		\cos^2\phi   	& 0 \\
		0   					& \left(\sqrt{1-\mu^2} \cos \phi - \mu \sin \phi\right)^2
	\end{array}
	\right].
\end{equation}
Accordingly, the tradeoff relation Eq.~\eqref{eq:CFIM tradeoff relation} holds in the considered measurement protocol through this classical FIM and, furthermore, is tight only under the condition as
\begin{equation}\label{eq:tight condition}
	\mu \cos \phi + \sqrt{1-\mu^2} \sin \phi \leq 0, 
\end{equation}
with $\phi \in [0,\pi]$ (see Appendix \ref{appd:condition} for details).
These results manifest that, for specific $\mu$, there exists an optimal $\phi$-phase range to reach the least estimation variances.

\begin{figure}[htbp]
	\includegraphics[width=0.85\columnwidth]{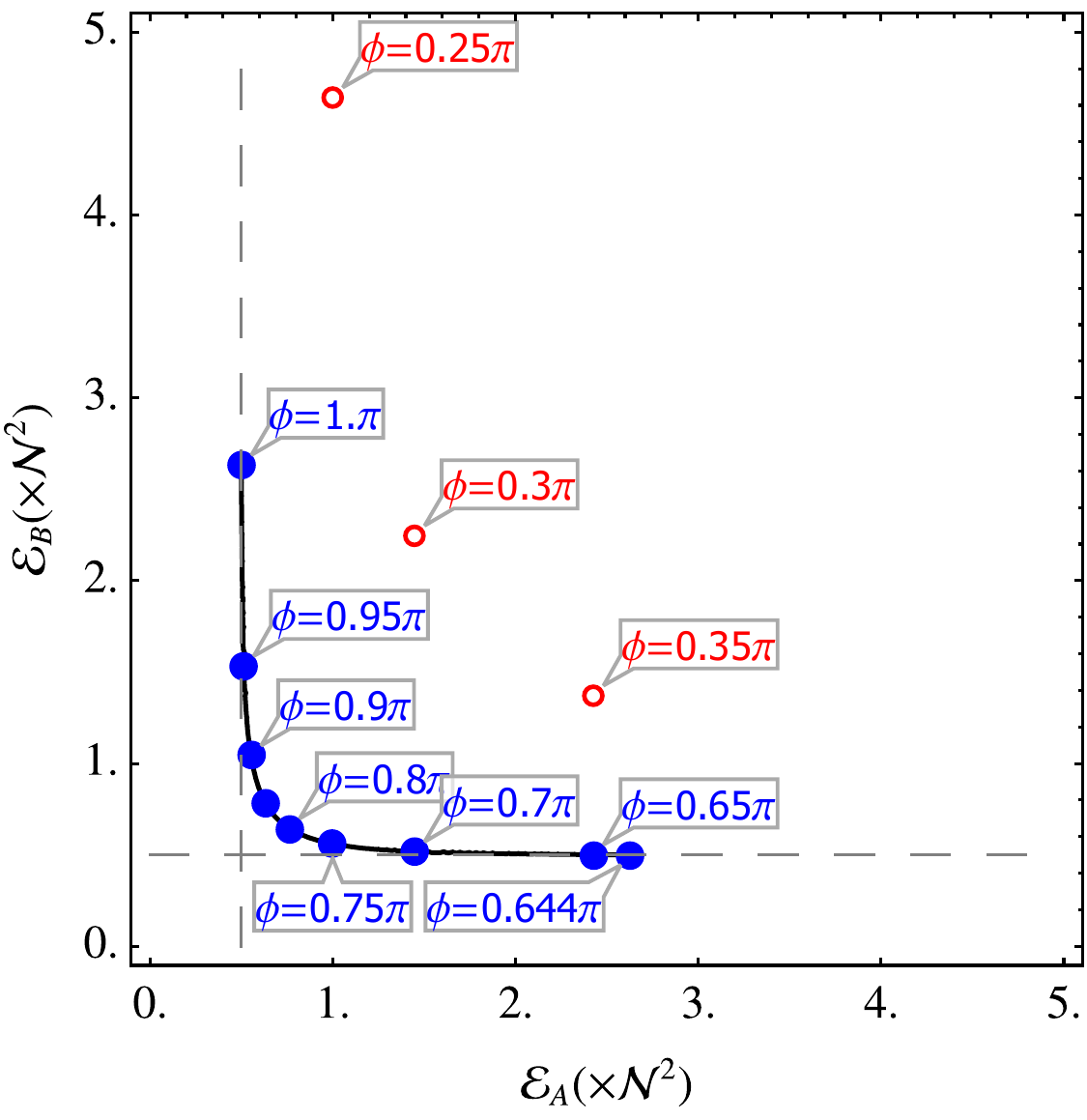}
	\centering
	\caption{
	The tradeoff relation (via black solid line) and the estimation errors of the measurement protocol \eqref{eq:measurement} (via dots with various $\phi$ attached).
	The phase $\phi$ is sampled at intervals of $0.5\pi$ and $\mu=0.9$ without loss of generality.
	However, for specific $\phi$-values that violate the condition \eqref{eq:tight condition}, the associated dots deviate too far from this region to be displayed. 
	The blue solid dots denote $\phi$ obeying the tight condition \eqref{eq:tight condition}, and exactly coincide with the line of the tradeoff relation.
	Conversely, the red open dots that denote $\phi$ violating the tight condition deviate away from the line.
	Besides, the gray dashed lines denote the quantum limits as $1/(2\mathcal{N}^2)$ for individual estimation of $A$ or $B$.
	}
	\label{fig:measure perform}
\end{figure}

Based on the classical FIM \eqref{eq:mat cF}, Fig.~\ref{fig:measure perform} exhibits several error bounds of the measurement protocol \eqref{eq:measurement} corresponding to various phases $\phi$ that  either satisfy or violate the condition \eqref{eq:tight condition}.
In this figure, we further compare the error bound of this measurement protocol and the tradeoff relation \eqref{eq:error tradeoff relation}.
It reveals that the errors of the considered measurements \eqref{eq:measurement} saturate the tradeoff bound when the measurement phase $\phi$ conforms to the condition \eqref{eq:tight condition}.
Our findings display that phase $\phi$ acts as an available parameter to adjust the individual estimation error.
In fact, as the phase $\phi$ increases, the estimation for parameter $A$ becomes more accurate at the expense of a rise in estimation error for parameter $B$.
In contrast, if the phase $\phi$ is adjusted to violate the condition \eqref{eq:tight condition}, as shown in Fig.~\ref{fig:measure perform}, the measurement protocol \eqref{eq:measurement} cannot attain the ultimate tradeoff bound.

The extension to the detection with squeezed-state input is straightforward.
Let $S(r)$ be the squeeze operator with $r \in \mathbb{R}$, and the quantum state for measurement is instead a two-mode displaced squeezed state
$\ket{\beta_1, \beta_2 ; r} 
:= D(\beta_1) S(r)\ket{0} \otimes  D(\beta_2) S(r)\ket{0} 
= S(r) D(\widetilde{\beta}_1) \ket{0} \otimes S(r) D(\widetilde{\beta}_2) \ket{0}$ with $\widetilde{\beta}_j = \beta_j \cosh r + \beta_j^\ast \sinh r$.
Due to the squeezing effect, the IRTR Eq.~\eqref{eq:CFIM tradeoff relation}, precision tradeoff relation Eq.~\eqref{eq:error tradeoff relation}, and  classical FIM Eq.~\eqref{eq:mat cF} are slightly modified by replacing Euclidean norm and incompatibility coefficient with 
\begin{align}\label{eq:replace}
   \mathcal{N} &\rightarrow
   \mathcal{N}_r = \sqrt{\frac{T}{2} \qty(e^{2r} |\chi_{xG}|^2 + e^{-2r} |\chi_{pG}|^2 )}, \notag \\
   \mu &\rightarrow
   \mu_r = \qty(\frac{\mathcal{N}}{\mathcal{N}_r})^2 \mu.
\end{align}
Consequently, the measurement protocol similar to Eq.~\eqref{eq:measurement} can be acquired, and the corresponding optimal $\phi$-phase range Eq.~\eqref{eq:tight condition} should be modified by the replacement $\mu \rightarrow \mu_r$ as Eq.~\eqref{eq:replace}.
This squeezing case is discussed in Appendix \ref{appd:sqz} in detail.

\section{Detuned GW sensors}

As shown in Fig.~\ref{fig:Comp 4 Miu}, while both estimation variances at $\mu = 0$ can simultaneously reach their respective quantum CRBs, such simultaneous attainability is impossible for growing $\mu$, owing to the tradeoff relation \eqref{eq:error tradeoff relation}.
Thus, $\mu$ is a key factor affecting the quantum limit of linear devices (see Fig.~\ref{fig:Miu Influence} in Appendix \ref{appd:fig equal weight}). 
As a significant example, detuned interferometers serve as sensitive linear devices to detect GW kilohertz signals from postmerger remnants of binary neutron-star mergers, thereby probing extreme matter \cite{Lasky_2015, BAIOTTI2019103714, PhysRevLett.116.061102, universe7040097, PhysRevD.79.044030, Ott2009, PhysRevX.4.041004, PhysRevLett.122.061102, Abbott2017, Abbott2019}.
However, based on the interferometer susceptibility $\vec{\chi}$ in Ref.~\cite{PhysRevLett.132.130801}, the nonzero detuned frequency $\Delta$ and signal frequency $\Omega$ at several kilohertz lead to the incompatibility coefficient as
$c_{12} = \mu = {2 \Delta \Omega}/{\left(\gamma^2 + \Delta^2 + \Omega^2\right)}$ with bandwidth $\gamma$.
Thus, the inevitable tradeoff relation \eqref{eq:error tradeoff relation} can guide researchers to learn the precisions of both estimated parameters of GW signals.
That is, one has to confront an either-or situation where only one of both parameters can access higher precision region by manipulating the tunable coefficient $\phi$ in the measurement \eqref{eq:measurement} and, simultaneously, the precision of the other one is inevitably sacrificed.

\begin{figure}[tbhp]
 \includegraphics[width=0.7\columnwidth]{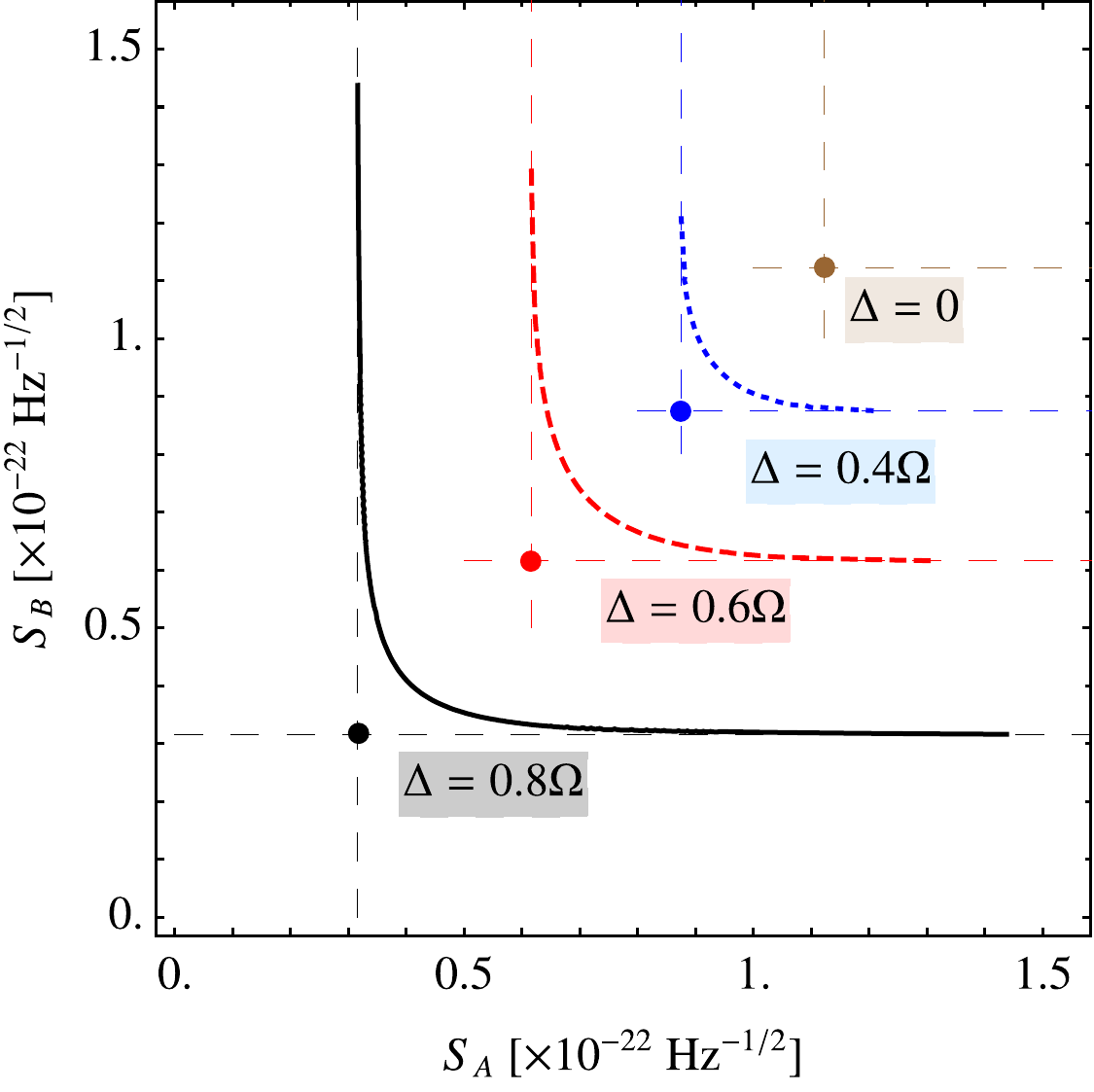}
	\centering
	\caption{The relation between both sensitivities $S_A$ and $S_B$ of two parameters of interest, $A$ and $B$, in the detuned GW interferometer for signal frequency $\Omega = 3~\mathrm{kHz}$ with $\Delta = 0.8 \Omega$ (black solid line), $\Delta = 0.6 \Omega$ (red dashed line), $\Delta = 0.4 \Omega$ (blue dotted line), and $\Delta = 0$ (brown dot). 
    The dashed straight lines corresponding to different detunings $\Delta$ denote the optimal sensitivities of the individual measurements for independent signal parameters $A$ and $B$.
	}
	\label{fig:Detuned Sens}
\end{figure}

For signal frequency $\Omega = 3~\mathrm{kHz}$, Fig.~\ref{fig:Detuned Sens} exhibits how the detuned frequency $\Delta$ influences the tradeoff relation between the two sensitivities $S_A:=\sqrt{\mathcal{E}_A T}$ and $S_B:=\sqrt{\mathcal{E}_B T}$ of two parameters of interest, $A$ and $B$, in a detuned configuration with arm length $L=4$~km, bandwidth $\gamma = 42$~Hz, circulating power $P=750$~kW, and carrier laser frequency $\omega_0=282$~THz \cite{PhysRevLett.132.130801}.
As illustrated, when the detuned frequency $\Delta$ approaches the signal frequency $\Omega$ ($\Delta \sim \Omega$), the signal sensing is capable of achieving smaller errors but encounters a tradeoff between the errors, hindering the achievement of individual quantum CRB.
Therefore, this work that establishes the precision tradeoff relation for linear measurements becomes available when detuned GW sensors are applied to detect kilohertz signals with enhanced precision.

\section{Conclusion}
In summary, we have thoroughly analyzed the estimation errors of the independent parameters of a monochromatic signal jointly detected in a linear quantum device. 
Based on the framework of information regret tradeoff relation, we have put forward an ultimate, analytical expression of the tradeoff relation between the attainable estimation variances.
Compared with the Holevo Cram\'er-Rao bound proposed in previous work \cite{PhysRevLett.132.130801}, our results arise from incompatible measurements governed by the Heisenberg uncertainty principle, and thus can clearly reveal the quantum limits on the estimation errors that exhibit an inherent tradeoff relation.
Moreover, we have found that the tradeoff effect turns stronger with larger incompatibility coefficient and can be treated as a strict criterion to assess the quantum multiparameter estimation.
Eventually, we have also proved that a proposed measurement can attain the tradeoff bound under a specific condition, and show that the errors are adjustable by shifting a phase of the proposed measurement.

This work provides a useful criterion for assessing interferometric gravitational-wave detectors operated in the detuned regime to enhance sensitivities of kilohertz signals.
The detuned frequency is beneficial to improving precision, but is directly related to the incompatibility coefficient, inevitably leading to our tradeoff bound.
Our future interest could lie in applying our findings to other systems, e.g., optical cavities \cite{Geraci2019}, transmon qubits \cite{Chen2023}, and atomic clocks \cite{Filzinger2025}, for dark matter search.

\vspace{0.5cm}
\section*{Acknowledgments}
This work is supported by the Quantum Science and Technology-National Science and Technology Major Project (Grant No. 2024ZD0301000), the National Natural Science Foundation of China (Grants No. 92476118 and No. 12275062), and Key R{\&}D Program of Zhejiang (Grant No. 2026C01004).

\bibliography{apsreflist_24GW}

@article{Luca2025,
  title = {Distributed Quantum Multiparameter Estimation with Optimal Local Measurements},
  author = {Pezz\`e, Luca and Smerzi, Augusto},
  journal = {Phys. Rev. Lett.},
  volume = {135},
  issue = {26},
  pages = {260805},
  numpages = {7},
  year = {2025},
  month = {Dec},
  publisher = {American Physical Society},
  doi = {10.1103/f2jf-bg7g},
  url = {https://link.aps.org/doi/10.1103/f2jf-bg7g}
}

@article{Filzinger2025,
  title     = {Ultralight Dark Matter Search with Space-Time Separated Atomic Clocks and Cavities},
  author    = {Filzinger, Melina and Caddell, Ashlee R. and Jani, Dhruv and Steinel, Martin and Giani, Leonardo and Huntemann, Nils and Roberts, Benjamin M.},
  journal   = {Phys. Rev. Lett.},
  volume    = {134},
  issue     = {3},
  pages     = {031001},
  numpages  = {8},
  year      = {2025},
  month     = {Jan},
  publisher = {American Physical Society},
  doi       = {10.1103/PhysRevLett.134.031001},
  url       = {https://link.aps.org/doi/10.1103/PhysRevLett.134.031001}
}

@article{Chen2023,
  title     = {Detecting Hidden Photon Dark Matter Using the Direct Excitation of Transmon Qubits},
  author    = {Chen, Shion and Fukuda, Hajime and Inada, Toshiaki and Moroi, Takeo and Nitta, Tatsumi and Sichanugrist, Thanaporn},
  journal   = {Phys. Rev. Lett.},
  volume    = {131},
  issue     = {21},
  pages     = {211001},
  numpages  = {8},
  year      = {2023},
  month     = {Nov},
  publisher = {American Physical Society},
  doi       = {10.1103/PhysRevLett.131.211001},
  url       = {https://link.aps.org/doi/10.1103/PhysRevLett.131.211001}
}

@article{Geraci2019,
  title     = {Searching for Ultralight Dark Matter with Optical Cavities},
  author    = {Geraci, Andrew A. and Bradley, Colin and Gao, Dongfeng and Weinstein, Jonathan and Derevianko, Andrei},
  journal   = {Phys. Rev. Lett.},
  volume    = {123},
  issue     = {3},
  pages     = {031304},
  numpages  = {6},
  year      = {2019},
  month     = {Jul},
  publisher = {American Physical Society},
  doi       = {10.1103/PhysRevLett.123.031304},
  url       = {https://link.aps.org/doi/10.1103/PhysRevLett.123.031304}
}

@article{Belliardo2021,
  doi       = {10.1088/1367-2630/ac04ca},
  url       = {https://dx.doi.org/10.1088/1367-2630/ac04ca},
  year      = {2021},
  month     = {jun},
  publisher = {IOP Publishing},
  volume    = {23},
  number    = {6},
  pages     = {063055},
  author    = {Belliardo, Federico and Giovannetti, Vittorio},
  title     = {Incompatibility in quantum parameter estimation},
  journal   = {New Journal of Physics}
}

@article{Yung2024,
  title     = {Comparison of estimation limits for quantum two-parameter estimation},
  author    = {Yung, Simon K. and Conlon, Lorc\'an O. and Zhao, Jie and Lam, Ping Koy and Assad, Syed M.},
  journal   = {Phys. Rev. Res.},
  volume    = {6},
  issue     = {3},
  pages     = {033315},
  numpages  = {12},
  year      = {2024},
  month     = {Sep},
  publisher = {American Physical Society},
  doi       = {10.1103/PhysRevResearch.6.033315},
  url       = {https://link.aps.org/doi/10.1103/PhysRevResearch.6.033315}
}

@article{Abbott2017,
  doi       = {10.3847/2041-8213/aa9a35},
  url       = {https://dx.doi.org/10.3847/2041-8213/aa9a35},
  year      = {2017},
  month     = {dec},
  publisher = {The American Astronomical Society},
  volume    = {851},
  number    = {1},
  pages     = {L16},
  author    = { B. P. Abbott et al.},
  title     = {Search for Post-merger Gravitational Waves from the Remnant of the Binary Neutron Star Merger GW170817},
  journal   = {The Astrophysical Journal Letters}
}

@article{Abbott2019,
  doi       = {10.3847/1538-4357/ab0f3d},
  url       = {https://dx.doi.org/10.3847/1538-4357/ab0f3d},
  year      = {2019},
  month     = {apr},
  publisher = {The American Astronomical Society},
  volume    = {875},
  number    = {2},
  pages     = {160},
  author    = {B. P. Abbott et al.},
  title     = {Search for Gravitational Waves from a Long-lived Remnant of the Binary Neutron Star Merger GW170817},
  journal   = {The Astrophysical Journal}
}

@inbook{Braginsky1992,
  author    = {Braginsky, V. B. and Khalilli, F.},
  title     = {Quantum Measurement},
  year      = {1992},
  publisher = {Cambridge University Press}
}

@article{PhysRevA.31.3761,
  title     = {Input and output in damped quantum systems: Quantum stochastic differential equations and the master equation},
  author    = {Gardiner, C. W. and Collett, M. J.},
  journal   = {Phys. Rev. A},
  volume    = {31},
  issue     = {6},
  pages     = {3761--3774},
  numpages  = {0},
  year      = {1985},
  month     = {Jun},
  publisher = {American Physical Society},
  doi       = {10.1103/PhysRevA.31.3761},
  url       = {https://link.aps.org/doi/10.1103/PhysRevA.31.3761}
}

@article{PhysRevA.95.012103,
  title     = {General quantum constraints on detector noise in continuous linear measurements},
  author    = {Miao, Haixing},
  journal   = {Phys. Rev. A},
  volume    = {95},
  issue     = {1},
  pages     = {012103},
  numpages  = {6},
  year      = {2017},
  month     = {Jan},
  publisher = {American Physical Society},
  doi       = {10.1103/PhysRevA.95.012103},
  url       = {https://link.aps.org/doi/10.1103/PhysRevA.95.012103}
}

@article{PhysRevD.65.042001,
  title     = {Signal recycled laser-interferometer gravitational-wave detectors as optical springs},
  author    = {Buonanno, Alessandra and Chen, Yanbei},
  journal   = {Phys. Rev. D},
  volume    = {65},
  issue     = {4},
  pages     = {042001},
  numpages  = {26},
  year      = {2002},
  month     = {Jan},
  publisher = {American Physical Society},
  doi       = {10.1103/PhysRevD.65.042001},
  url       = {https://link.aps.org/doi/10.1103/PhysRevD.65.042001}
}

@article{Kubo1966,
  doi       = {10.1088/0034-4885/29/1/306},
  url       = {https://dx.doi.org/10.1088/0034-4885/29/1/306},
  year      = {1966},
  month     = {jan},
  publisher = {},
  volume    = {29},
  number    = {1},
  pages     = {255},
  author    = {R Kubo},
  title     = {The fluctuation-dissipation theorem},
  journal   = {Reports on Progress in Physics}
}

@article{Li22JOSAB39,
  author    = {Guolong Li and Xiao-Ming Lu and Xiaoguang Wang and Jun Xin and Xingmin Li},
  journal   = {J. Opt. Soc. Am. B},
  number    = {1},
  pages     = {98--106},
  publisher = {Optica Publishing Group},
  title     = {Optomechanical gyroscope simultaneously estimating the position of the rotation axis},
  volume    = {39},
  month     = {Jan},
  year      = {2022},
  url       = {https://opg.optica.org/josab/abstract.cfm?URI=josab-39-1-98},
  doi       = {10.1364/JOSAB.441232}
}

@article{PhysRevLett.132.153602,
  title     = {Amplifying Frequency Up-Converted Infrared Signals with a Molecular Optomechanical Cavity},
  author    = {Zou, Fen and Du, Lei and Li, Yong and Dong, Hui},
  journal   = {Phys. Rev. Lett.},
  volume    = {132},
  issue     = {15},
  pages     = {153602},
  numpages  = {6},
  year      = {2024},
  month     = {Apr},
  publisher = {American Physical Society},
  doi       = {10.1103/PhysRevLett.132.153602},
  url       = {https://link.aps.org/doi/10.1103/PhysRevLett.132.153602}
}

@article{PhysRevLett.132.101001,
  title     = {GALILEO: Galactic Axion Laser Interferometer Leveraging Electro-Optics},
  author    = {Ebadi, Reza and Kaplan, David E. and Rajendran, Surjeet and Walsworth, Ronald L.},
  journal   = {Phys. Rev. Lett.},
  volume    = {132},
  issue     = {10},
  pages     = {101001},
  numpages  = {8},
  year      = {2024},
  month     = {Mar},
  publisher = {American Physical Society},
  doi       = {10.1103/PhysRevLett.132.101001},
  url       = {https://link.aps.org/doi/10.1103/PhysRevLett.132.101001}
}

@article{RevModPhys.86.121,
  title     = {Gravitational radiation detection with laser interferometry},
  author    = {Adhikari, Rana X.},
  journal   = {Rev. Mod. Phys.},
  volume    = {86},
  issue     = {1},
  pages     = {121--151},
  numpages  = {0},
  year      = {2014},
  month     = {Feb},
  publisher = {American Physical Society},
  doi       = {10.1103/RevModPhys.86.121},
  url       = {https://link.aps.org/doi/10.1103/RevModPhys.86.121}
}

@article{Danilishin12,
  title   = {Quantum Measurement Theory in Gravitational-Wave Detectors},
  author  = {Danilishin, Stefan L. and Khalili, Farid Ya.},
  journal = {Living Rev. Relativ.},
  volume  = {15},
  pages   = {5},
  year    = {2012},
  doi     = {10.doi.org/10.12942/lrr-2012-5},
  url     = {https://doi.org/10.12942/lrr-2012-5}
}

@article{PhysRevLett.132.130801,
  title     = {Achieving the Fundamental Quantum Limit of Linear Waveform Estimation},
  author    = {Gardner, James W. and Gefen, Tuvia and Haine, Simon A. and Hope, Joseph J. and Chen, Yanbei},
  journal   = {Phys. Rev. Lett.},
  volume    = {132},
  issue     = {13},
  pages     = {130801},
  numpages  = {7},
  year      = {2024},
  month     = {Mar},
  publisher = {American Physical Society},
  doi       = {10.1103/PhysRevLett.132.130801},
  url       = {https://link.aps.org/doi/10.1103/PhysRevLett.132.130801}
}

@inbook{Nagaoka2005,
  author    = { Hiroshi   Nagaoka },
  title     = {A New Approach to {Cramér-Rao} Bounds for Quantum State Estimation},
  booktitle = {Asymptotic Theory of Quantum Statistical Inference},
  year      = {2005},
  chapter   = {},
  pages     = {100-112},
  doi       = {10.1142/9789812563071_0009}
}

@article{Matsumoto2002,
  doi       = {10.1088/0305-4470/35/13/307},
  url       = {https://dx.doi.org/10.1088/0305-4470/35/13/307},
  year      = {2002},
  month     = {mar},
  publisher = {},
  volume    = {35},
  number    = {13},
  pages     = {3111},
  author    = {K Matsumoto},
  title     = {A new approach to the
               {Cramér-Rao}-type bound of the pure-state model},
  journal   = {Journal of Physics A: Mathematical and General}
}

@article{PhysRevA.61.042312,
  title     = {State estimation for large ensembles},
  author    = {Gill, Richard D. and Massar, Serge},
  journal   = {Phys. Rev. A},
  volume    = {61},
  issue     = {4},
  pages     = {042312},
  numpages  = {16},
  year      = {2000},
  month     = {Mar},
  publisher = {American Physical Society},
  doi       = {10.1103/PhysRevA.61.042312},
  url       = {https://link.aps.org/doi/10.1103/PhysRevA.61.042312}
}

@article{CAROLLO20201,
  title    = {Geometry of quantum phase transitions},
  journal  = {Physics Reports},
  volume   = {838},
  pages    = {1-72},
  year     = {2020},
  note     = {Geometry of quantum phase transitions},
  issn     = {0370-1573},
  doi      = {https://doi.org/10.1016/j.physrep.2019.11.002},
  url      = {https://www.sciencedirect.com/science/article/pii/S0370157319303655},
  author   = {Angelo Carollo and Davide Valenti and Bernardo Spagnolo}
}

@article{Kull2020,
  doi       = {10.1088/1751-8121/ab7f67},
  url       = {https://dx.doi.org/10.1088/1751-8121/ab7f67},
  year      = {2020},
  month     = {may},
  publisher = {IOP Publishing},
  volume    = {53},
  number    = {24},
  pages     = {244001},
  author    = {Ilya Kull and Philippe Allard Guérin and Frank Verstraete},
  title     = {Uncertainty and trade-offs in quantum multiparameter estimation},
  journal   = {J. Phys. A: Math. Theor.}
}

@article{Suzuki2020,
  title     = {Quantum state estimation with nuisance parameters},
  volume    = {53},
  issn      = {1751-8121},
  doi       = {10.1088/1751-8121/ab8b78},
  number    = {45},
  journal   = {J. Phys. A: Math. Theor.},
  publisher = {IOP Publishing},
  author    = {Suzuki, Jun and Yang, Yuxiang and Hayashi, Masahito},
  year      = {2020},
  month     = oct,
  pages     = {453001}
}

@article{Suzuki2019,
  author         = {Suzuki, Jun},
  title          = {Information Geometrical Characterization of Quantum Statistical Models in Quantum Estimation Theory},
  journal        = {Entropy},
  volume         = {21},
  year           = {2019},
  number         = {7},
  article-number = {703},
  url            = {https://www.mdpi.com/1099-4300/21/7/703},
  pubmedid       = {33267417},
  issn           = {1099-4300},
  doi            = {10.3390/e21070703}
}

@article{Suzuki2016,
  author   = {Suzuki, Jun},
  title    = {{Explicit formula for the Holevo bound for two-parameter qubit-state estimation problem}},
  journal  = {Journal of Mathematical Physics},
  volume   = {57},
  number   = {4},
  pages    = {042201},
  year     = {2016},
  month    = {04},
  issn     = {0022-2488},
  doi      = {10.1063/1.4945086},
  url      = {https://doi.org/10.1063/1.4945086}
}

@article{PhysRevLett.120.030404,
  title     = {Universally Fisher-Symmetric Informationally Complete Measurements},
  author    = {Zhu, Huangjun and Hayashi, Masahito},
  journal   = {Phys. Rev. Lett.},
  volume    = {120},
  issue     = {3},
  pages     = {030404},
  numpages  = {6},
  year      = {2018},
  month     = {Jan},
  publisher = {American Physical Society},
  doi       = {10.1103/PhysRevLett.120.030404},
  url       = {https://link.aps.org/doi/10.1103/PhysRevLett.120.030404}
}

@article{PhysRevLett.116.180402,
  title     = {Fisher-Symmetric Informationally Complete Measurements for Pure States},
  author    = {Li, Nan and Ferrie, Christopher and Gross, Jonathan A. and Kalev, Amir and Caves, Carlton M.},
  journal   = {Phys. Rev. Lett.},
  volume    = {116},
  issue     = {18},
  pages     = {180402},
  numpages  = {5},
  year      = {2016},
  month     = {May},
  publisher = {American Physical Society},
  doi       = {10.1103/PhysRevLett.116.180402},
  url       = {https://link.aps.org/doi/10.1103/PhysRevLett.116.180402}
}

@article{PhysRevA.94.052108,
  title     = {Compatibility in multiparameter quantum metrology},
  author    = {Ragy, Sammy and Jarzyna, Marcin and Demkowicz-Dobrza\ifmmode \acute{n}\else \'{n}\fi{}ski, Rafa\l{}},
  journal   = {Phys. Rev. A},
  volume    = {94},
  issue     = {5},
  pages     = {052108},
  numpages  = {11},
  year      = {2016},
  month     = {Nov},
  publisher = {American Physical Society},
  doi       = {10.1103/PhysRevA.94.052108},
  url       = {https://link.aps.org/doi/10.1103/PhysRevA.94.052108}
}

@article{Sidhu2021,
  title     = {Tight Bounds on the Simultaneous Estimation of Incompatible Parameters},
  author    = {Sidhu, Jasminder S. and Ouyang, Yingkai and Campbell, Earl T. and Kok, Pieter},
  year      = {2021},
  month     = feb,
  journal   = {Phys. Rev. X},
  volume    = {11},
  number    = {1},
  pages     = {011028},
  publisher = {American Physical Society},
  doi       = {10.1103/PhysRevX.11.011028},
}

@article{PhysRevA.101.022303,
  title     = {Generalized-mean {Cram\'er-Rao} bounds for multiparameter quantum metrology},
  author    = {Lu, Xiao-Ming and Ma, Zhihao and Zhang, Chengjie},
  journal   = {Phys. Rev. A},
  volume    = {101},
  issue     = {2},
  pages     = {022303},
  numpages  = {9},
  year      = {2020},
  month     = {Feb},
  publisher = {American Physical Society},
  doi       = {10.1103/PhysRevA.101.022303},
  url       = {https://link.aps.org/doi/10.1103/PhysRevA.101.022303}
}

@article{Sidhu2020,
  title     = {Geometric perspective on quantum parameter estimation},
  author    = {Sidhu, Jasminder S. and Kok, Pieter},
  journal   = {AVS Quantum Science},
  volume    = {2},
  issue     = {},
  pages     = {014701 },
  numpages  = {},
  year      = {2020},
  month     = {},
  publisher = {},
  doi       = {1010.1116/1.5119961},
  url       = {https://doi.org/10.1116/1.5119961}
}

@misc{albarelli2020upper,
  title         = {Upper bounds on the Holevo {Cram\'er-Rao} bound for multiparameter quantum parametric and semiparametric estimation},
  author        = {Francesco Albarelli and Mankei Tsang and Animesh Datta},
  year          = {2020},
  eprint        = {1911.11036},
  archiveprefix = {arXiv},
  primaryclass  = {quant-ph}
}

@misc{tsang2021holevo,
  title         = {The {Holevo Cram\'er-Rao} bound is at most thrice the {Helstrom} version},
  author        = {Mankei Tsang},
  year          = {2021},
  eprint        = {1911.08359},
  archiveprefix = {arXiv},
  primaryclass  = {quant-ph}
}

@article{PhysRevLett.123.200503,
  title     = {Evaluating the Holevo Cram\'er-Rao Bound for Multiparameter Quantum Metrology},
  author    = {Albarelli, Francesco and Friel, Jamie F. and Datta, Animesh},
  journal   = {Phys. Rev. Lett.},
  volume    = {123},
  issue     = {20},
  pages     = {200503},
  numpages  = {7},
  year      = {2019},
  month     = {Nov},
  publisher = {American Physical Society},
  doi       = {10.1103/PhysRevLett.123.200503},
  url       = {https://link.aps.org/doi/10.1103/PhysRevLett.123.200503}
}

@article{Rubio2018,
  doi       = {10.1088/2399-6528/aaa234},
  url       = {https://dx.doi.org/10.1088/2399-6528/aaa234},
  year      = {2018},
  month     = {jan},
  publisher = {IOP Publishing},
  volume    = {2},
  number    = {1},
  pages     = {015027},
  author    = {Jesús Rubio and Paul Knott and Jacob Dunningham},
  title     = {Non-asymptotic analysis of quantum metrology protocols beyond the Cramér–Rao bound},
  journal   = {Journal of Physics Communications}
}

@article{Carollo2019,
  doi       = {10.1088/1742-5468/ab3ccb},
  url       = {https://dx.doi.org/10.1088/1742-5468/ab3ccb},
  year      = {2019},
  month     = {sep},
  publisher = {IOP Publishing and SISSA},
  volume    = {2019},
  number    = {9},
  pages     = {094010},
  author    = {Angelo Carollo and Bernardo Spagnolo and Alexander A Dubkov and Davide Valenti},
  title     = {On quantumness in multi-parameter quantum estimation},
  journal   = {Journal of Statistical Mechanics: Theory and Experiment}
}

@article{BUSCH2007155,
  title    = {Heisenberg's uncertainty principle},
  journal  = {Physics Reports},
  volume   = {452},
  number   = {6},
  pages    = {155-176},
  year     = {2007},
  issn     = {0370-1573},
  doi      = {https://doi.org/10.1016/j.physrep.2007.05.006},
  url      = {https://www.sciencedirect.com/science/article/pii/S0370157307003481},
  author   = {Paul Busch and Teiko Heinonen and Pekka Lahti}
}

@article{Fujiwara2006,
  doi       = {10.1088/0305-4470/39/40/014},
  url       = {https://dx.doi.org/10.1088/0305-4470/39/40/014},
  year      = {2006},
  month     = {sep},
  publisher = {},
  volume    = {39},
  number    = {40},
  pages     = {12489},
  author    = {Akio Fujiwara},
  title     = {Strong consistency and asymptotic efficiency for adaptive quantum estimation problems},
  journal   = {Journal of Physics A: Mathematical and General}
}

@article{BRAUNSTEIN1996135,
  title    = {Generalized Uncertainty Relations: Theory, Examples, and Lorentz Invariance},
  journal  = {Annals of Physics},
  volume   = {247},
  number   = {1},
  pages    = {135-173},
  year     = {1996},
  issn     = {0003-4916},
  doi      = {https://doi.org/10.1006/aphy.1996.0040},
  url      = {https://www.sciencedirect.com/science/article/pii/S0003491696900408},
  author   = {Samuel L. Braunstein and Carlton M. Caves and G.J. Milburn}
}

@inbook{LehmannBook1998,
  author    = {E. Lehmann and G. Casella},
  booktitle = {Theory of Point Estimation},
  year      = {1998},
  publisher = {Springer-Verlag},
  address   = {New York}
}

@inbook{CasellaBook2002,
  author    = {G. Casella and R. L. Berger},
  booktitle = {Statistical Inference, 2nd ed.},
  year      = {2002},
  publisher = {Duxbury Press},
  address   = {Pacific Grove}
}

@inbook{WassermanBook2010,
  author    = {L. Wasserman},
  booktitle = {All of Statistics: A Concise Course in Statistical Inference},
  year      = {2010},
  publisher = {Springer Publishing Company},
  address   = {Incorporated}
}

@inbook{KayBook1993,
  author    = {S. M. Kay},
  booktitle = {Fundamentals of Statistical Signal Processing, Volume I: Estimation Theory},
  year      = {1993},
  publisher = {Prentice Hall},
  address   = {}
}

@article{Fisher1925,
  title   = {Theory of Statistical Estimation},
  volume  = {22},
  doi     = {10.1017/S0305004100009580},
  number  = {5},
  journal = {Mathematical Proceedings of the Cambridge Philosophical Society},
  author  = {Fisher, R. A.},
  year    = {1925},
  pages   = {700--725}
}

@article{Fisher1922,
  issn      = {02643952},
  url       = {http://www.jstor.org/stable/91208},
  author    = {R. A. Fisher},
  journal   = {Philosophical Transactions of the Royal Society of London. Series A, Containing Papers of a Mathematical or Physical Character},
  number    = {},
  pages     = {309--368},
  publisher = {The Royal Society},
  title     = {On the Mathematical Foundations of Theoretical Statistics},
  urldate   = {2024-03-18},
  volume    = {222},
  year      = {1922}
}

@article{PhysRevX.10.031023,
  title     = {Quantum Semiparametric Estimation},
  author    = {Tsang, Mankei and Albarelli, Francesco and Datta, Animesh},
  journal   = {Phys. Rev. X},
  volume    = {10},
  issue     = {3},
  pages     = {031023},
  numpages  = {28},
  year      = {2020},
  month     = {Jul},
  publisher = {American Physical Society},
  doi       = {10.1103/PhysRevX.10.031023},
  url       = {https://link.aps.org/doi/10.1103/PhysRevX.10.031023}
}

@inbook{HayashiBook2005,
  editor    = {M. Hayashi},
  booktitle = {Asymptotic theory of quantum statistical inference: Selected Papers},
  year      = {2005},
  publisher = {World Scientific Publishing Company},
  address   = {}
}

@article{Personick1054643,
  author   = {Personick, S.},
  journal  = {IEEE Transactions on Information Theory},
  title    = {Application of quantum estimation theory to analog communication over quantum channels},
  year     = {1971},
  volume   = {17},
  number   = {3},
  pages    = {240-246},
  doi      = {10.1109/TIT.1971.1054643}
}

@inbook{HolevoBook1982,
  author    = {A. S. Holevo},
  booktitle = {Probabilistic and Statistical Aspects of Quantum Theory},
  year      = {1982},
  publisher = {North-Holland},
  address   = {Amsterdam}
}

@inbook{HelstromBook1976,
  author    = {C. W. Helstrom},
  booktitle = {Quantum Detection and Estimation Theory},
  year      = {1976},
  publisher = {Academic Press},
  address   = {New York}
}

@article{Belavkin1976,
  author   = {Belavkin, V. P.},
  journal  = {Theoretical and Mathematical Physics},
  title    = {Generalized uncertainty relations and efficient measurements in quantum systems},
  year     = {1976},
  volume   = {26},
  number   = {},
  pages    = {213-222},
  doi      = {10.1007/BF01032091},
  url      = {https://doi.org/10.1007/BF01032091}
}

@article{Yuen1055103,
  author   = {Yuen, H. and Lax, M.},
  journal  = {IEEE Transactions on Information Theory},
  title    = {Multiple-parameter quantum estimation and measurement of nonselfadjoint observables},
  year     = {1973},
  volume   = {19},
  number   = {6},
  pages    = {740-750},
  doi      = {10.1109/TIT.1973.1055103}
}

@article{Helstrom1054108,
  author   = {Helstrom, C.},
  journal  = {IEEE Transactions on Information Theory},
  title    = {The minimum variance of estimates in quantum signal detection},
  year     = {1968},
  volume   = {14},
  number   = {2},
  pages    = {234-242},
  doi      = {10.1109/TIT.1968.1054108}
}

@article{HELSTROM1967101,
  title    = {Minimum mean-squared error of estimates in quantum statistics},
  journal  = {Physics Letters A},
  volume   = {25},
  number   = {2},
  pages    = {101-102},
  year     = {1967},
  issn     = {0375-9601},
  doi      = {https://doi.org/10.1016/0375-9601(67)90366-0},
  url      = {https://www.sciencedirect.com/science/article/pii/0375960167903660},
  author   = {C.W. Helstrom}
}

@article{10.1093/nsr/nwx029,
  author   = {Cai, Rong-Gen and Cao, Zhoujian and Guo, Zong-Kuan and Wang, Shao-Jiang and Yang, Tao},
  title    = {{The gravitational-wave physics}},
  journal  = {National Science Review},
  volume   = {4},
  number   = {5},
  pages    = {687-706},
  year     = {2017},
  month    = {04},
  issn     = {2095-5138},
  doi      = {10.1093/nsr/nwx029},
  url      = {https://doi.org/10.1093/nsr/nwx029}
}

@article{Lasky_2015,
  title   = {Gravitational Waves from Neutron Stars: A Review},
  volume  = {32},
  doi     = {10.1017/pasa.2015.35},
  journal = {Publications of the Astronomical Society of Australia},
  author  = {Lasky, Paul D.},
  year    = {2015},
  pages   = {e034}
}

@article{Ott2009,
  title     = {The gravitational-wave signature of core-collapse supernovae},
  author    = {Ott, C. D.},
  journal   = {Class. Quantum Grav.},
  volume    = {26 },
  pages     = {063001},
  year      = {2009},
  publisher = {IOP},
  doi       = {10.1088/0264-9381/26/6/063001},
  url       = {https://iopscience.iop.org/article/10.1088/0264-9381/26/6/063001}
}

@article{PhysRevX.4.041004,
  title     = {Source Redshifts from Gravitational-Wave Observations of Binary Neutron Star Mergers},
  author    = {Messenger, C. and Takami, Kentaro and Gossan, Sarah and Rezzolla, Luciano and Sathyaprakash, B. S.},
  journal   = {Phys. Rev. X},
  volume    = {4},
  issue     = {4},
  pages     = {041004},
  numpages  = {12},
  year      = {2014},
  month     = {Oct},
  publisher = {American Physical Society},
  doi       = {10.1103/PhysRevX.4.041004},
  url       = {https://link.aps.org/doi/10.1103/PhysRevX.4.041004}
}

@article{PhysRevD.79.044030,
  title     = {Gravitational waves from black hole-neutron star binaries: Classification of waveforms},
  author    = {Shibata, Masaru and Kyutoku, Koutarou and Yamamoto, Tetsuro and Taniguchi, Keisuke},
  journal   = {Phys. Rev. D},
  volume    = {79},
  issue     = {4},
  pages     = {044030},
  numpages  = {27},
  year      = {2009},
  month     = {Feb},
  publisher = {American Physical Society},
  doi       = {10.1103/PhysRevD.79.044030},
  url       = {https://link.aps.org/doi/10.1103/PhysRevD.79.044030}
}

@article{universe7040097,
  author         = {Andersson, Nils},
  title          = {A Gravitational-Wave Perspective on Neutron-Star Seismology},
  journal        = {Universe},
  volume         = {7},
  year           = {2021},
  number         = {4},
  pages          = {97},
  url            = {https://www.mdpi.com/2218-1997/7/4/97},
  issn           = {2218-1997},
  doi            = {10.3390/universe7040097}
}

@article{PhysRevLett.116.061102,
  title         = {Observation of Gravitational Waves from a Binary Black Hole Merger},
  author        = {Abbott, B. P. and Abbott, R. and Abbott, T. D. and {\it et al.} },
  collaboration = {LIGO Scientific Collaboration and Virgo Collaboration},
  journal       = {Phys. Rev. Lett.},
  volume        = {116},
  issue         = {6},
  pages         = {061102},
  numpages      = {16},
  year          = {2016},
  month         = {Feb},
  publisher     = {American Physical Society},
  doi           = {10.1103/PhysRevLett.116.061102},
  url           = {https://link.aps.org/doi/10.1103/PhysRevLett.116.061102}
}

@article{AdvancedLIGO,
  title         = {Advanced LIGO},
  author        = {Aasi, J and Abbott, B. P. and Abbott, R. and Abbott, T. and {\it et al.} },
  collaboration = {LIGO Scientific Collaboration},
  journal       = {Class. Quantum Grav.},
  volume        = {32},
  pages         = {074001},
  year          = {2015},
  month         = {Mar},
  publisher     = {IOPscience},
  doi           = {10.1088/0264-9381/32/7/074001},
  url           = {https://iopscience.iop.org/article/10.1088/0264-9381/32/7/074001}
}

@article{KAGRA,
  title     = {KAGRA: 2.5 generation interferometric gravitational wave detector},
  volume    = {3},
  issn      = {2397-3366},
  url       = {http://dx.doi.org/10.1038/s41550-018-0658-y},
  doi       = {10.1038/s41550-018-0658-y},
  number    = {1},
  journal   = {Nature Astronomy},
  publisher = {Springer Science and Business Media LLC},
  author    = {Akutsu, T. and Ando, M. and Arai, K. and {\it et al.} },
  year      = {2019},
  month     = jan,
  pages     = {35--40}
}

@article{LIGOVirgoKAGRA,
  title     = {Prospects for observing and localizing gravitational-wave transients with Advanced LIGO, Advanced Virgo and KAGRA},
  volume    = {23},
  url       = {https://doi.org/10.1007/s41114-020-00026-9},
  doi       = {https://doi.org/10.1007/s41114-020-00026-9},
  number    = {3},
  journal   = {Living Rev Relativ},
  publisher = {Springer Science and Business Media LLC},
  author    = {Abbott, B.P. and Abbott, R. and Abbott, T.D. and {\it et al.} },
  year      = {2020},
  month     = Sept,
  pages     = {1--69}
}

@article{BAIOTTI2019103714,
  title    = {Gravitational waves from neutron star mergers and their relation to the nuclear equation of state},
  journal  = {Progress in Particle and Nuclear Physics},
  volume   = {109},
  pages    = {103714},
  year     = {2019},
  issn     = {0146-6410},
  doi      = {https://doi.org/10.1016/j.ppnp.2019.103714},
  url      = {https://www.sciencedirect.com/science/article/pii/S0146641019300493},
  author   = {Luca Baiotti}
}

@article{PhysRevLett.122.061102,
  title     = {Identifying a First-Order Phase Transition in Neutron-Star Mergers through Gravitational Waves},
  author    = {Bauswein, Andreas and Bastian, Niels-Uwe F. and Blaschke, David B. and Chatziioannou, Katerina and Clark, James A. and Fischer, Tobias and Oertel, Micaela},
  journal   = {Phys. Rev. Lett.},
  volume    = {122},
  issue     = {6},
  pages     = {061102},
  numpages  = {8},
  year      = {2019},
  month     = {Feb},
  publisher = {American Physical Society},
  doi       = {10.1103/PhysRevLett.122.061102},
  url       = {https://link.aps.org/doi/10.1103/PhysRevLett.122.061102}
}

@article{PhysRevX.9.031040,
  title         = {GWTC-1: A Gravitational-Wave Transient Catalog of Compact Binary Mergers Observed by LIGO and Virgo during the First and Second Observing Runs},
  author        = {Abbott, B. P. and Abbott, R. and Abbott, T. D. and {\it et al} },
  collaboration = {LIGO Scientific Collaboration and Virgo Collaboration},
  journal       = {Phys. Rev. X},
  volume        = {9},
  issue         = {3},
  pages         = {031040},
  numpages      = {49},
  year          = {2019},
  month         = {Sep},
  publisher     = {American Physical Society},
  doi           = {10.1103/PhysRevX.9.031040},
  url           = {https://link.aps.org/doi/10.1103/PhysRevX.9.031040}
}

@article{PhysRevX.11.021053,
  title         = {GWTC-2: Compact Binary Coalescences Observed by LIGO and Virgo during the First Half of the Third Observing Run},
  author        = {Abbott, R. and Abbott, T. D. and Abraham, S. and  {\it et al} },
  collaboration = {LIGO Scientific Collaboration and Virgo Collaboration},
  journal       = {Phys. Rev. X},
  volume        = {11},
  issue         = {2},
  pages         = {021053},
  numpages      = {52},
  year          = {2021},
  month         = {Jun},
  publisher     = {American Physical Society},
  doi           = {10.1103/PhysRevX.11.021053},
  url           = {https://link.aps.org/doi/10.1103/PhysRevX.11.021053}
}

@article{science.abc7397,
  author   = {Salvatore Vitale },
  title    = {The first 5 years of gravitational-wave astrophysics},
  journal  = {Science},
  volume   = {372},
  number   = {6546},
  pages    = {eabc7397},
  year     = {2021},
  doi      = {10.1126/science.abc7397},
  url      = {https://www.science.org/doi/abs/10.1126/science.abc7397}
}

@article{Miller2019,
  author  = {Miller, M. C.  and Yunes, N.},
  title   = {The new frontier of gravitational waves},
  journal = {Nature},
  volume  = {568},
  pages   = {469-676},
  year    = {2019},
  doi     = {https://doi.org/10.1038/s41586-019-1129-z},
  url     = {https://www.nature.com/articles/s41586-019-1129-z#citeas}
}

@article{AdvancedVirgo,
  author  = {Acernese, F.  and Agathos, M. and Agatsuma, K. and  {\it et al} },
  title   = {Advanced Virgo: a second-generation interferometric gravitational wave detector},
  journal = {Class. Quantum Grav.},
  volume  = {32},
  pages   = {024001},
  year    = {2015},
  doi     = {10.1088/0264-9381/32/2/024001},
  url     = {https://iopscience.iop.org/article/10.1088/0264-9381/32/2/024001}
}

@misc{gardner2023holevo,
  title         = {Holevo Cram\'er-Rao Bound for waveform estimation of gravitational waves},
  author        = {James W. Gardner and Tuvia Gefen and Simon A. Haine and Joseph J. Hope and Yanbei Chen},
  year          = {2023},
  eprint        = {2308.06253},
  archiveprefix = {arXiv},
  primaryclass  = {gr-qc}
}

@article{PhysRevLett.126.120503,
  title     = {Incorporating {Heisenberg}'s Uncertainty Principle into Quantum Multiparameter Estimation},
  author    = {Lu, Xiao-Ming and Wang, Xiaoguang},
  journal   = {Phys. Rev. Lett.},
  volume    = {126},
  issue     = {12},
  pages     = {120503},
  numpages  = {7},
  year      = {2021},
  month     = {Mar},
  publisher = {American Physical Society},
  doi       = {10.1103/PhysRevLett.126.120503},
  url       = {https://link.aps.org/doi/10.1103/PhysRevLett.126.120503}
}

@inbook{Rao1992,
  author    = {Rao, C. Radhakrishna},
  editor    = {Kotz, Samuel
               and Johnson, Norman L.},
  title     = {Information and the Accuracy Attainable in the Estimation of Statistical Parameters},
  booktitle = {Breakthroughs in Statistics: Foundations and Basic Theory},
  year      = {1992},
  publisher = {Springer New York},
  address   = {New York, NY},
  pages     = {235--247},
  isbn      = {978-1-4612-0919-5},
  doi       = {10.1007/978-1-4612-0919-5_16},
  url       = {https://doi.org/10.1007/978-1-4612-0919-5_16}
}

@inbook{Cramer1946,
  author    = {Harald Cram{\`e}r},
  booktitle = {Mathematical Methods of Statistics},
  year      = {1946},
  publisher = {Princeton University Press},
  address   = {Princeton},
  isbn      = {9781400883868}
}

@article{PhysRevLett.119.050801,
  title     = {Towards the Fundamental Quantum Limit of Linear Measurements of Classical Signals},
  author    = {Miao, Haixing and Adhikari, Rana X and Ma, Yiqiu and Pang, Belinda and Chen, Yanbei},
  journal   = {Phys. Rev. Lett.},
  volume    = {119},
  issue     = {5},
  pages     = {050801},
  numpages  = {6},
  year      = {2017},
  month     = {Aug},
  publisher = {American Physical Society},
  doi       = {10.1103/PhysRevLett.119.050801},
  url       = {https://link.aps.org/doi/10.1103/PhysRevLett.119.050801}
}

@article{PhysRevLett.72.3439,
  title     = {Statistical distance and the geometry of quantum states},
  author    = {Braunstein, Samuel L. and Caves, Carlton M.},
  journal   = {Phys. Rev. Lett.},
  volume    = {72},
  issue     = {22},
  pages     = {3439--3443},
  numpages  = {0},
  year      = {1994},
  month     = {May},
  publisher = {American Physical Society},
  doi       = {10.1103/PhysRevLett.72.3439},
  url       = {https://link.aps.org/doi/10.1103/PhysRevLett.72.3439}
}

@article{Liu_2020,
  doi       = {10.1088/1751-8121/ab5d4d},
  url       = {https://dx.doi.org/10.1088/1751-8121/ab5d4d},
  year      = {2019},
  month     = {dec},
  publisher = {IOP Publishing},
  volume    = {53},
  number    = {2},
  pages     = {023001},
  author    = {Jing Liu and Haidong Yuan and Xiao-Ming Lu and Xiaoguang Wang},
  title     = {Quantum Fisher information matrix and multiparameter estimation},
  journal   = {J. Phys. A: Math. Theor.},
}

@article{Xia2023,
  title   = {Toward incompatible quantum limits on multiparameter estimation},
  author  = {Binke Xia and Jingzheng Huang and Hongjing Li and Han Wang and Guihua Zeng},
  journal = {Nat. Commun.},
  volume  = {14},
  pages   = {1021},
  year    = {2023},
  doi     = {https://doi.org/10.1038/s41467-023-36661-3},
  url     = {https://doi.org/10.1038/s41467-023-36661-3}
}

@article{Tsang2011,
  title   = {Fundamental Quantum Limit to Waveform Estimation},
  author  = {Tsang, Mankei and Wiseman, Howard M. and Caves, Carlton M.},
  year    = {2011},
  month   = mar,
  journal = {Phys. Rev. Lett.},
  volume  = {106},
  number  = {9},
  pages   = {090401},
  doi     = {10.1103/PhysRevLett.106.090401},
}

\onecolumngrid
\newpage

\setcounter{equation}{0}
\setcounter{figure}{0}
\setcounter{table}{0}
\setcounter{page}{1}
\makeatletter
\renewcommand{\thefigure}{S\arabic{figure}}
\renewcommand{\thetable}{S\arabic{table}}

\appendix
\counterwithin{subsection}{section}

\section{System and Symplectic Transformation}\label{appd:sys}
\subsection{system}
In the system governed by Eq.~\eqref{eq:linear system}, the detected optical field is described by the quadratures in frequency-domain as
\begin{align}
    x (\Omega) &= \int_{-\infty}^\infty
    \dd{t} \frac{1}{\sqrt{2}}
    [a(t) + a^\dagger(t)] 
    e^{i \Omega t}, \notag \\
    p (\Omega) &= \int_{-\infty}^\infty
    \dd{t} \frac{-i}{\sqrt{2}}
    [a(t) - a^\dagger(t)] 
    e^{i \Omega t},
\end{align}
with the corresponding annihilation (creation) operator $a$ ($a^\dagger$).
The real and imaginary parts of $x(\Omega)$ and $p(\Omega)$ obey the commutation relation as
$[ \Re[x(\Omega)], \Re[p(\Omega')] ] = [ \Im[x(\Omega)], \Im[p(\Omega')] ] = i \pi \delta (\Omega - \Omega') \sim i {T}/{2}$.
The last result is obtained because the function $\delta (\Omega)$ is the Dirac delta distribution in the frequency domain with units of time such that $\delta (0) \sim T/2\pi$ with integration time $T$.

The real and imaginary parts of $x(\Omega)$ and $p(\Omega)$ constitute a Hermitian-operator vector as
$\mathbf {x}=(x_1, p_1, x_2, p_2) := \sqrt{\frac{2}{T} } (\Re[x(\Omega)], \Re[p(\Omega)], \Im[x(\Omega)], \Im[p(\Omega)])$.
Then the parameters $A$ and $B$ are encoded in the elements of ${\mathbf x}$ as
\begin{align}\label{eq:vec x with unit vector}
    {\mathbf x}=
     {{\mathbf x}}^{(0)} + A \mathbf{d}_A + B \mathbf{d}_B,
\end{align}
where the first equation is explicitly expressed as 
\begin{equation}\label{eq:expand vec x with unit vector}
	\mqty [
    x_1 \\
    p_1 \\
    x_2 \\
    p_2
    ] 
    = \mqty [
    x_1^{(0)} \\
    p_1^{(0)} \\
    x_2^{(0)} \\
    p_2^{(0)}
    ] 
    + \sqrt{\frac{T}{2}} \mqty[
	\Re [\chi_{xG}] \\
	\Re [\chi_{pG}] \\
	\Im [\chi_{xG}] \\
	\Im [\chi_{pG}]
	] A
    + \sqrt{\frac{T}{2}} \mqty[
	-\Im [\chi_{xG}] \\
	  -\Im [\chi_{pG}] \\
    \Re [\chi_{xG}] \\
	\Re [\chi_{pG}]
	] B.
\end{equation}
Hence, the above definition makes the elements of the operator vector $\mathbf{x}$ obey the canonical commutation relation as 
$[x_j ,p_k]  = i \delta_{jk}$
and $[x_j ,x_k] = [p_j ,p_k] = 0$.
It means that this linear system is equivalent to two independent harmonic oscillators.
According to the simple correspondence, 
\begin{align}\label{eq:vec ds}
    \mathbf{d}_A
    &=\sqrt{\frac{T}{2}} 
    (\Re [\chi_{xG}], \Re [\chi_{pG}], \Im [\chi_{xG}], \Im [\chi_{pG}])^\mathrm{T}, \notag \\
    \mathbf{d}_B
    &= \sqrt{\frac{T}{2}} ( -\Im [\chi_{xG}], -\Im [\chi_{pG}], \Re [\chi_{xG}], \Re [\chi_{pG}])^\mathrm{T}
\end{align}
are two orthogonal vectors with a common Euclidean norm 
$\mathcal{N} = \sqrt{ \frac{T}{2} \left(\left| \chi_{xG} \right|^2 + \left| \chi_{pG} \right|^2\right)}$, 
and the quadrature vector 
${\mathbf x}^{(0)} := \qty(x_1^{(0)}, p_1^{(0)}, x_2^{(0)}, p_2^{(0)})  $ 
represents the initial state without signal impacting.

Obviously, the parameters of interest $A$ and $B$ displace two independent harmonic oscillators with canonical quadrature $(x_j, p_j)$ ($j=1,2$).
Moreover, the na{\"i}ve optimal unbiased estimates
$\hat{A}_\mathrm{nv} = (\mathbf{d}_A \cdot \mathbf{x})/\mathcal{N}^2$ and 
$\hat{B}_\mathrm{nv} = ( \mathbf{d}_B \cdot \mathbf{x}) /\mathcal{N}^2 $ of $A$ and $B$ satisfy 
\begin{align}\label{eq:naive commu in SM}
  \qty[ \hat{A}_\mathrm{nv},  \hat{B}_\mathrm{nv}] 
  = \frac{T}{2 \mathcal{N}^4} 
  & \Bigg[ \Re [\chi_{xG}] x_1^{(0)} 
  + \Re [\chi_{pG}] p_1^{(0)}
  + \Im [\chi_{xG}] x_2^{(0)} 
  + \Im [\chi_{pG}] p_2^{(0)}, 
  \notag \\
  &-\Im [\chi_{xG}] x_1^{(0)} 
  - \Im [\chi_{pG}] p_1^{(0)}
  + \Re [\chi_{xG}] x_2^{(0)} 
  + \Re [\chi_{pG}] p_2^{(0)}
  \Bigg]  
  = \frac{i \mu}{\mathcal{N}^2}  
\end{align}
with 
$\mu := T \mathcal{N}^{-2} 
(
\Re\left[ \chi_{pG}  \right] 
\Im\left[ \chi_{xG}  \right]   
- \Re\left[ \chi_{xG} \right] \Im \left[ \chi_{pG}  \right]
)$ ($0\leq \mu \leq 1$).

\subsection{symplectic transformation}
After a transformation of a 4-by-4 orthogonal symplectic matrix $\mathcal{M}$ for observable vector $\mathbf{x}$ in Eq.~\eqref{eq:vec x with unit vector} or \eqref{eq:vec x}, the vacuum fluctuation for the transformed vector
$\mathbf{X} = \mathcal{M} \mathbf{x}$ is unchanged, that is, the 4-by-4 covariance matrix remains
$\mathrm{diag}(\frac{1}{2}, \frac{1}{2},\frac{1}{2},\frac{1}{2})^\mathrm{T}$.
Moreover, the observables in the transformed vector $\mathbf{X}$ obey the same commutation relation.
Nevertheless, there are still four degrees of freedom to simplify the form of the vector $\mathbf{X}$.

An appropriate symplectic transformation $\mathcal{M}$ can be chosen to transform the observable vector ${\mathbf x}$ into the other one 
${\mathbf X} := ({X}_1, {P}_1, {X}_2, {P}_2)^\mathrm{T}$ 
as~\cite{PhysRevLett.132.130801}
\begin{equation}\label{eq:vec X}
    {\mathbf X} 
    = \mathcal{M} \mathbf{x} 
    = {\mathbf X}^{(0)} + A \mathcal{N}
    \left[\begin{array}{c}
         1  \\
         0  \\
         0  \\
         0
    \end{array}\right]  + B \mathcal{N}
    \left[\begin{array}{c}
         0               \\
         \mu             \\
         \sqrt{1-\mu^2}  \\
         0
    \end{array}\right] 
    = {\mathbf X}^{(0)} + A'
    \mathbf{n}_A  + B'
    \mathbf{n}_B ,
\end{equation}
where two rescaled parameters, $A' := \mathcal{N} A$ and $B' := \mathcal{N} B$, and two unit vectors, $\mathbf{n}_A = \mathcal{M} \mathbf{d}_A / \mathcal{N} = (1, 0, 0, 0 )^\mathrm{T}$ and $\mathbf{n}_B = \mathcal{M} \mathbf{d}_B / \mathcal{N} = (0, \mu, \sqrt{1-\mu^2}, 0)^\mathrm{T}$, are defined for compact derivations in the following.
The elements in ${{\mathbf X}}$, ${X}_j$ and ${P}_j$ with $j=1,2$, still obey $[{X}_j, {P}_k] = i \delta_{jk}$ and $[X_j ,X_k] = [P_j ,P_k] = 0$ based on the symplectic transformation $\mathcal{M}$, and thus the vector ${{\mathbf X}}$ also contains the amplitude and phase quadratures for two independent harmonic oscillators, just as the vector $\mathbf{x}$.
Obviously, the na{\"i}ve optimal unbiased estimates 
$\hat{A}_\mathrm{nv} = (\mathbf{d}_A \cdot \mathbf{x})/\mathcal{N}^2$ and 
$\hat{B}_\mathrm{nv} =( \mathbf{d}_B \cdot \mathbf{x}) /\mathcal{N}^2 $
can be rewritten as
\begin{align}
    \hat{A}_\mathrm{nv} &= \frac{\mathbf{d}_A \cdot \mathbf{x}}{\mathcal{N}^2}
    = \frac{\mathbf{d}_A \mathcal{M}^\mathrm{T}}{\mathcal{N}} \cdot 
    \frac{\mathcal{M} \mathbf{x} }{ \mathcal{N}}
    = \frac{\mathbf{n_A} \cdot \mathbf{X}}{\mathcal{N}}
    = \frac{X_1}{\mathcal{N}}, \notag \\
    \hat{B}_\mathrm{nv} &= \frac{\mathbf{d}_B \cdot \mathbf{x}}{\mathcal{N}^2}
    = \frac{\mathbf{d}_B \mathcal{M}^\mathrm{T}}{\mathcal{N}} \cdot 
    \frac{\mathcal{M} \mathbf{x} }{ \mathcal{N}}
    = \frac{\mathbf{n_B} \cdot \mathbf{X}}{\mathcal{N}}
    =\frac{\mu P_1 + \sqrt{1-\mu^2} X_2}{\mathcal{N}},
\end{align}
still obeying the commutation relation Eq~\eqref{eq:naive commu} or \eqref{eq:naive commu in SM} due to the orthogonal symplectic matrix $\mathcal{M}$.

If the initial state in the linear device, reflected by ${{\mathbf x}}^{(0)}$ and ${{\mathbf X}}^{(0)}$, is prepared in a vacuum state, the signal with the estimated parameters $A$ and $B$ displaces the vacuum state due to Eq.~\eqref{eq:vec X}.
Based on Eq.~\eqref{eq:vec X}, the parameters of interest $A$ and $B$, as well as $A' = \mathcal{N} A$ and $B' = \mathcal{N} B$, are thus encoded in a two-mode coherent state 
$\ket{\psi} = \ket{\alpha_1,\alpha_2} $ with
\begin{align}\label{eq:alphas}
	\alpha_1 &=
    \frac{\mathcal{N}}{\sqrt{2}} \left(A + i \mu B \right) =
        \frac{1}{\sqrt{2}} \left(A' + i \mu B' \right), 
        \notag \\
	\alpha_2 &= 
    \frac{\mathcal{N}}{\sqrt{2}} \sqrt{1-\mu^2} B =
        \frac{1}{\sqrt{2}} \sqrt{1-\mu^2} B',
\end{align} 
in the Schr\"{o}dinger picture.
This treatment is highly available of assessing sensing precision in linear measurement devices.

\section{The Derivation of Quantum Geometric Tensor $\mathcal{Q}$}\label{appd:mat Q}

\subsection{method 1}
Based on Eq.~\eqref{eq:vec x}, if the initial state, reflected by $\mathbf{x}^{(0)}$, is a vacuum state, the parameters $A$ and $B$ are thus encoded in a two-mode coherent state 
$\ket{\psi} = \ket{\beta_1,\beta_2}$ where the real and imaginary parts of 
$\beta_j = \beta_j^\mathrm{r} + i \beta_j^\mathrm{i}$, i.e., $\beta_j^\mathrm{r} := \Re\beta_j$ and $\beta_j^\mathrm{i} := \Im \beta_j$ can be obtained from Eq.~\eqref{eq:beta} as
\begin{align}\label{eq:trans para to beta}
	\mqty [
    \beta_1^\mathrm{r} \\
    \beta_1^\mathrm{i} \\
    \beta_2^\mathrm{r} \\
    \beta_2^\mathrm{i}
    ] 
    =\frac{\sqrt{T}}{2} \mqty[
     \Re[\chi_{xG}] A - \Im[\chi_{xG}] B   \\
    \Re[\chi_{pG}] A -\Im[\chi_{pG}] B \\
     \Im[\chi_{xG}] A + \Re[\chi_{xG}] B  \\
    \Im[\chi_{pG}] A +  \Re[\chi_{pG}] B 
    ]
    = \frac{\sqrt{T}}{2}
	\mqty[
	\Re[\chi_{xG}] & - \Im[\chi_{xG}] \\
	\Re[\chi_{pG}] & - \Im[\chi_{pG}] \\
	\Im[\chi_{xG}] &   \Re[\chi_{xG}] \\
	\Im[\chi_{pG}] &   \Re[\chi_{pG}]
	]
    \mqty [A \\ B]
    = \mathcal{P} \mqty [A \\ B],
\end{align}
with a transformation matrix as
\begin{equation}
    \mathcal{P}
    =\frac{\sqrt{T}}{2}
	\mqty[
	\Re[\chi_{xG}] & - \Im[\chi_{xG}] \\
	\Re[\chi_{pG}] & - \Im[\chi_{pG}] \\
	\Im[\chi_{xG}] &   \Re[\chi_{xG}] \\
	\Im[\chi_{pG}] &   \Re[\chi_{pG}]
	].
\end{equation}
The quantum geometric tensor of two-mode coherent state $\ket{\beta_1, \beta_2}$ with respect to the parameters
$\beta_1^\mathrm{r}:= \Re \beta_1$, $\beta_1^\mathrm{i}:= \Im \beta_1$, $\beta_2^\mathrm{r}:= \Re \beta_2$, and $\beta_2^\mathrm{i}:= \Im \beta_2$, 
i.e., the real and imaginary parts of $\beta_j$ ($j=1,2$), is known as \cite{PhysRevLett.126.120503}
\begin{equation}\label{eq:Q for ReIm part}
    \mathcal{Q}_0
    =4 \mqty[
	1 & i &  0 & 0 \\
   -i & 1 &  0 & 0 \\
	  0 & 0 &  1 & i \\
	  0 & 0 & -i & 1
	].
\end{equation}
We thus obtain the quantum geometric tensor with respect to $A$ and $B$ as
\begin{align}
    \mathcal{Q} = \mathcal{P}^\mathrm{T} \mathcal{Q}_0 \mathcal{P} &= T
    \mqty[
	|\chi_{xG}|^2 + |\chi_{pG}|^2 & 2i (\Re \chi_{pG} \Im \chi_{xG} - \Re \chi_{xG} \Im \chi_{pG}) \\
   -2 i (\Re \chi_{pG} \Im \chi_{xG} - \Re \chi_{xG} \Im \chi_{pG}) & |\chi_{xG}|^2 + |\chi_{pG}|^2  
	] \notag \\
    &= 2 \mathcal{N}^2 
    \mqty(
	1      & i \mu \\
   - i \mu & 1  
	).
\end{align}
Here we use the common Euclidean norm $\mathcal{N} = |\mathbf{d}_A| = |\mathbf{d}_B|$ and the coefficient $\mu$ defined in Eq.~\eqref{eq:naive commu}.

\subsection{method 2}
Here we use the two-mode coherent state $\ket{\alpha_1, \alpha_2}$ with amplitudes as Eq.~\eqref{eq:alphas} to
derive the quantum geometric tensor $\mathcal{Q}$ in Eq.~\eqref{eq:mat Q}.
If the real and imaginary parts of $\alpha$ are $\vartheta_1 \equiv \Re \alpha$ and $\vartheta_2 \equiv \Im \alpha$, for a coherent state $\ket{\alpha}$ with an complex amplitude $\alpha$, the coherent state $\ket{\alpha}$ obeys
\begin{equation}\label{eq:derivation to ReIm}
	\frac{\partial \ket{\alpha}}{\partial \vartheta_1}
	=(-\vartheta_1 + a^\dagger) \ket{\alpha}, \qquad
	\frac{\partial \ket{\alpha}}{\partial \vartheta_2}
	=(-\vartheta_2+ i a^\dagger) \ket{\alpha}.
\end{equation}
For Brevity, the derivations in this part are expressed by the rescaled parameters $A'$ and $B'$.
In the end, the result for the parameters of interest $A = A'/ \mathcal{N}$ and $B = B' / \mathcal{N}$ can be easily obtained.

Based on the two-mode coherent state $\ket{\psi} = \ket{\alpha_1, \alpha_2}$ with both amplitudes in Eq.~\eqref{eq:alphas}, we define their real and imaginary parts as
\begin{align}\label{eq:ReIm of alpha}
 \vartheta_{11} & \equiv \Re \alpha_1 = \frac{A'}{\sqrt{2}},
 \qquad
 \vartheta_{12} \equiv \Im \alpha_1 = \frac{ \mu B'}{\sqrt{2}}, \notag \\
 \vartheta_{21} & \equiv \Re \alpha_2 = \sqrt{\frac{1 - \mu^2}{2}}  B',
 \qquad
 \vartheta_{22} \equiv \Im \alpha_2 = 0.
\end{align}
Applying Eq.~\eqref{eq:derivation to ReIm} to the two-mode coherent state $\ket{\psi} = \ket{\alpha_1, \alpha_2}$, we get
\begin{align}\label{eq:derivation to para}
	\frac{\partial \ket{\psi}}{\partial A'} 	
	&= \frac{\partial \vartheta_{11}}{\partial A'} \frac{\partial \ket{\alpha_1, \alpha_2}}{\partial \vartheta_{11}}
	= \frac{1}{\sqrt{2}} 
	\left(- \frac{A'}{\sqrt{2}} + a_1^\dagger \right) \ket{\psi}, \notag \\
	\frac{\partial \ket{\psi}}{\partial B'}
	&= \frac{\partial \vartheta_{12}}{\partial B'} \frac{\partial \ket{\alpha_1, \alpha_2}}{\partial \vartheta_{12}}
	+ \frac{\partial \vartheta_{21}}{\partial B'} \frac{\partial \ket{\alpha_1, \alpha_2}}{\partial \vartheta_{21}}
	= \left[\frac{\mu}{\sqrt{2}} \left( - \frac{\mu B'}{\sqrt{2}} + i a_1^\dagger \right) +  \sqrt{\frac{1-\mu^2}{2}}
	\left(- \sqrt{\frac{1-\mu^2}{2}} B' + a_2^\dagger \right)\right]
	\ket{\psi} \notag \\
	&= \frac{1}{\sqrt{2}} \left(-\frac{ B'}{\sqrt{2}} + i \mu a_1^\dagger + \sqrt{1-\mu^2}  a_2^\dagger\right) \ket{\psi}.
\end{align}
As for the pure state $\ket{\psi}$, the quantum geometric tensor is given by
\begin{equation}\label{eq:mat Q for PureState}
	\mathcal{Q} _{jk}
	= 4 \left( \frac{\partial \bra{\psi}}{\partial \theta_j} \right)
	(\mathds{1} - \ket{\psi}\bra{\psi})
	\left( \frac{\partial \ket{\psi}}{\partial \theta_k} \right).
\end{equation}
Thus, using Eqs.~\eqref{eq:derivation to para}, we obtain the results as
\begin{align}
	\left( \frac{\partial \bra{\psi}}{\partial A'} \right) \left( \frac{\partial \ket{\psi}}{\partial A'} \right)
	&= \frac{1}{2} 
	\left(\left| - \frac{A'}{\sqrt{2}} + \alpha_1 \right|^2 +1 \right), 
	\qquad
	\left( \frac{\partial \bra{\psi}}{\partial A'} \right)  \ket{\psi}\bra{\psi} \left( \frac{\partial \ket{\psi}}{\partial A'} \right)
	= \frac{1}{2} 
	\left| - \frac{A'}{\sqrt{2}} + \alpha_1 \right|^2, \notag \\
	\left( \frac{\partial \bra{\psi}}{\partial B'} \right) \left( \frac{\partial \ket{\psi}}{\partial B'} \right)
	&= \frac{1}{2}
	\left(\left| - \frac{B'}{\sqrt{2}} - i \mu \alpha_1 + \sqrt{1-\mu^2} \alpha_2 \right|^2 +1 \right), \notag \\
	\left( \frac{\partial \bra{\psi}}{\partial B'} \right)  \ket{\psi}\bra{\psi} \left( \frac{\partial \ket{\psi}}{\partial B'} \right)
	&= \frac{1}{2} 
	\left| - \frac{ B'}{\sqrt{2}} - i \mu \alpha_1 + \sqrt{1-\mu^2} \alpha_2 \right|^2, \notag \\
	\left( \frac{\partial \bra{\psi}}{\partial A'} \right) \left( \frac{\partial \ket{\psi}}{\partial B'} \right)
	&= \frac{1}{2}
	\left[\left(- \frac{A'}{\sqrt{2}} + \alpha_1 \right)
	\left(-\frac{B'}{\sqrt{2}} + i \mu \alpha_1^\ast + \sqrt{1-\mu^2}  \alpha_2^\ast\right) + i \mu \right], \notag \\
	\left( \frac{\partial \bra{\psi}}{\partial A'} \right)  \ket{\psi}\bra{\psi} \left( \frac{\partial \ket{\psi}}{\partial B'} \right)
	&= \frac{1}{2}
	\left(- \frac{ A'}{\sqrt{2}} + \alpha_1 \right)
	\left(-\frac{ B'}{\sqrt{2}} + i \mu \alpha_1^\ast + \sqrt{1-\mu^2}  \alpha_2^\ast\right),
\end{align}
and put them into Eq.~\eqref{eq:mat Q for PureState} to acquire the quantum geometric tensor with respect to $A'$ and $B'$ as
\begin{equation}
	2 \left(
	\begin{array}{cc}
		1       & i\mu \\
		-i\mu & 1
	\end{array}
	\right).
\end{equation}
Eventually, the above tensor yields the quantum geometric tensor with respect to $A$ and $B$ as
\begin{equation}
	\mathcal{Q}=2 \mathcal{N}^2 \left(
	\begin{array}{cc}
		1       & i\mu \\
		-i\mu & 1
	\end{array}
	\right),
\end{equation}
the same with the final result in method 1.

\section{The Degree of the Tradeoff Relation via Equal Weight Case $w=0.5$}\label{appd:fig equal weight}
\begin{figure}[tbhp]
\includegraphics[width=0.4\columnwidth]{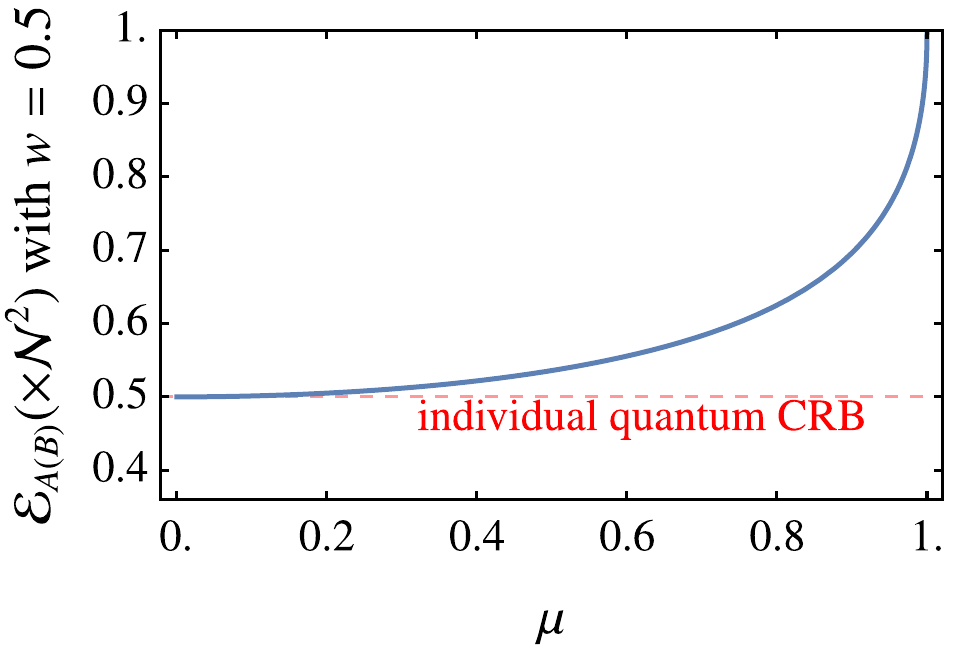}
	\centering
	\caption{
		Estimation variance with equal weight, i.e., $\mathcal{E}_{A(B)}$ with $w=0.5$, versus $\mu$.
        The red dashed straight line represents $\mathcal{E}_{A(B)} = 0.5$ as the individual quantum CRB.
	}
	\label{fig:Miu Influence}
\end{figure}

We use the estimation variances with equal weight, i.e., $\mathcal{E}_{A(B)}$ with $w=0.5$, to reflect the gap between allowed bound and individual quantum CRB, and depict its variation with $\mu$ in Fig.~\ref{fig:Miu Influence}.
The monotone increase manifests that the tradeoff effect becomes obvious when $\mu$ is large in a specific device.

\section{The Derivation of Classical Fisher Information}

Based on Eq.~\eqref{eq:vec X}, the compatible unbiased estimates for parameters of interest $A$ and $B$ are given by
\begin{align}\label{eq:measurement in SM}
	\hat{A} = \mathcal{N}^{-1}
    [{X}_1 - \mathcal{T}(\phi) {P}_2],
    \quad
	\hat{B} = \mathcal{N}^{-1}
    [- \mathcal{S}(\phi) {P}_1 +  \mathcal{C}(\phi) {X}_2],
\end{align}
with coefficients
\begin{align}
	\mathcal{C} (\phi) &= \cos \phi / \left( \sqrt{1-\mu^2} \cos \phi - \mu \sin \phi \right), \notag \\
	\mathcal{S} (\phi) &= \sin \phi / \left(\sqrt{1-\mu^2} \cos \phi - \mu \sin \phi\right), \notag \\
    \mathcal{T} (\phi) &= \mathcal{S} (\phi) / \mathcal{C} (\phi) =  \tan \phi,
\end{align} 
i.e., Eq.~\eqref{eq:measurement} in the main text.
Since the norm $\mathcal{N}$ in Eq.~\eqref{eq:measurement} or \eqref{eq:measurement in SM} is too bothersome to obtain the result clearly, we use the corresponding unbiased estimates for scaled parameters $A'$ and $B'$ as
\begin{align}\label{eq:reduced measurement}
	\hat{A}' = \mathcal{N} \hat{A}
    = {X}_1 - \mathcal{T}(\phi) {P}_2 ,
    \quad
	\hat{B}' = \mathcal{N} \hat{B}
    = - \mathcal{S}(\phi) {P}_1 +  \mathcal{C}(\phi) {X}_2.
\end{align}
In the end, the result for the parameters of interest $A = A'/ \mathcal{N}$ and $B = B' / \mathcal{N}$ can be easily obtained.

In the following, we provide the optimal observables in the {\bf Schr\"{o}dinger picture} to obtain classical Fisher information matrix (FIM).
Here $\hat{\mathbf{x}} = (\hat{x}_1, \hat{p}_1, \hat{x}_2, \hat{p}_2)$ and $\hat{\mathbf{X}} = (\hat{X}_1, \hat{P}_1, \hat{X}_2, \hat{P}_2)$ are introduced to represent the quadratures in the Schr\"{o}dinger picture, corresponding to $\mathbf{x}$ and $\mathbf{X}$ respectively.
First, according to \eqref{eq:vec X}, the quadratures $X_j$ and $P_j$ are displacements from vacuum quadratures $X_j^{(0)}$ and $P_j^{(0)}$, and thus are equivalent to the case where the quadratures $\hat{X}_j$ and $\hat{P}_j$ in the Schr\"{o}dinger picture apply to the two-mode coherent state $\ket{\alpha_1, \alpha_2}$.
The corresponding unbiased estimates for scaled parameters $A'$ and $B'$ in Eq. ~\eqref{eq:reduced measurement}, as well as the ones for $A$ and $B$ in Eq.~\eqref{eq:measurement} or \eqref{eq:measurement in SM}, shift into the Schr\"{o}dinger picture via directly replacing the quadratures $X_j$ and $P_j$ with $\hat{X}_j$ and $\hat{P}_j$ in the Schr\"{o}dinger picture as
\begin{align}\label{eq:reduced measurement in Schr}
	\hat{A}' =
     \hat{X}_1 - \mathcal{T}(\phi) \hat{P}_2 ,
    \quad
	\hat{B}' = 
     - \mathcal{S}(\phi) \hat{P}_1 +  \mathcal{C}(\phi) \hat{X}_2.
\end{align}
Then, in Eq.~\eqref{eq:vec X}, the linear relation $\mathbf{X} = \mathcal{M} \mathbf{x}$ between the quadrature vectors $\mathbf{x}$ and $\mathbf{X}$ is transferred into the form in the Schr\"{o}dinger picture as
\begin{equation}\label{eq:Omega transfer}
    \hat{\mathbf{X}} = \Omega^{-1} \mathcal{M}^\mathrm{T} \Omega \hat{\mathbf{x}}
    = \Omega^{T} \mathcal{M}^\mathrm{T} \Omega \hat{\mathbf{x}}, 
\end{equation}
with a matrix $\Omega = \mqty(0 & -1 \\ 1 & 0) \bigoplus \mqty(0 & -1 \\ 1 & 0)$, owing to the displacement operator $D(\beta_j) = \exp [i\sqrt{2} (\beta_j^\mathrm{i} \hat{x}_j - \beta_j^\mathrm{r} \hat{p}_j) ] = \exp[i \sqrt{2}  \mqty( \beta_j^\mathrm{r} & \beta_j^\mathrm{i}) \mqty(0 & -1 \\ 1 & 0) \mqty( \hat{x}_j \\ \hat{p}_j) ]$
for two-mode coherent state
$\ket{\beta_1, \beta_2} = D(\beta_1) \otimes D(\beta_2) \ket{0,0}$.

\subsection{simultaneous eigenstate}\label{appd:eigenstate}

In  the Schr\"{o}dinger picture, denote by $\xi$ and $\eta$ the eigenvalues of $\hat{A}'$ and $\hat{B}'$ respectively, and the corresponding simultaneous eigenstate is
\begin{equation}\label{eq:simultaneous eigenstate}
	\ket{\xi, \eta}
	=\frac{1}{\sqrt{2\pi \mathcal{T} \mathcal{S}}}
	\int e^{-i\frac{\eta}{\mathcal{S}} x} 
	\ket{x}_{X_1} \otimes \ket{\frac{x-\xi}{\mathcal{T}} }_{P_2} \dd x,
\end{equation}
where  $\ket{x}_{X_1} $ and $\ket{(x-\xi)/{\mathcal{T}} }_{P_2}$ are eigenstates of ${X}_1$ and ${P}_2$, with the eigenvalues $x$ and $(x-\xi)/\mathcal{T}$, respectively.

Here we prove that the state \eqref{eq:simultaneous eigenstate} is a simultaneous eigenstate of the observables $\hat{A}'$ and $\hat{B}'$ in Eq.~\eqref{eq:reduced measurement in Schr}.
On the one hand,
\begin{equation}
	\hat{A}' \ket{\xi, \eta} = [\hat{X}_1 - \mathcal{T} (\phi) \hat{P}_2] \ket{\xi, \eta}
	=\left(x - \mathcal{T} \cdot \frac{x-\xi}{\mathcal{T}} \right) \ket{\xi, \eta}
	=\xi \ket{\xi, \eta},
\end{equation}
On the other hand, this state can be rewritten as
\begin{equation}
	\begin{split}
		\ket{\xi, \eta}
		&= \frac{1}{\sqrt{2\pi \mathcal{T} \mathcal{S}}}
		\int \dd{p} \ket{p}_{P_1} \bra{p}  
		\int \dd{x'} \ket{x'}_{X_2} \bra{x'}
		\int e^{-i\frac{\eta}{\mathcal{S}} x} 
		\ket{x}_{X_1} \otimes \ket{\frac{x-\xi}{\mathcal{T}} }_{P_2} \dd{x} \\
		&= \frac{1}{\sqrt{2\pi \mathcal{T} \mathcal{S}}} 
		\iiint \ket{p}_{P_1} \otimes \ket{x'}_{X_2}
		e^{-i\frac{\eta}{\mathcal{S}} x} 
	\left(\bra{p}_{P_1} \cdot \ket{x}_{X_1} \right)
		\left(\bra{x'}_{X_2} \cdot \ket{\frac{x-\xi}{\mathcal{T}} }_{P_2} \right) 
        \dd{p} \dd{x'} \dd{x} \\
		&= \frac{1}{\sqrt{2\pi \mathcal{T} \mathcal{S}}} 
		\iiint \ket{p}_{P_1} \otimes \ket{x'}_{X_2}
		\left(e^{-i\frac{\eta}{\mathcal{S}} x}\right) 
		\left(\frac{1}{\sqrt{2\pi}} e^{-i p x} \right)
		\left(\frac{1}{\sqrt{2\pi}} e^{i x' \frac{x-\xi}{\mathcal{T}} } \right) \dd{p} \dd{x'} \dd{x} \\
		&= \frac{1}{\sqrt{2\pi \mathcal{T} \mathcal{S}}} 
		\iiint \ket{p}_{P_1} \otimes \ket{x'}_{X_2}
		\left(\frac{1}{2\pi} e^{i\left(\frac{x'}{\mathcal{T}} -p -\frac{\eta}{\mathcal{S}} \right) x}\right) 
		\left( e^{-i \frac{\xi}{\mathcal{T}}  x' } \right) \dd{p} \dd{x'} \dd{x} \\
		&= \frac{1}{\sqrt{2\pi \mathcal{T} \mathcal{S}}} 
		\iint \ket{p}_{P_1} \otimes \ket{x'}_{X_2}
		\delta\left( \frac{x'}{\mathcal{T}} -p -\frac{\eta}{\mathcal{S}} \right)
		\left( e^{-i \frac{\xi}{\mathcal{T}}  x' } \right) \dd{p} \dd{x'} \\
		&= \frac{1}{\sqrt{2\pi \mathcal{T} \mathcal{S}}} 
		\iint \ket{p}_{P_1} \otimes \ket{x'}_{X_2}
		\mathcal{T} \delta\left[x' - \mathcal{T} \left(p +\frac{\eta}{\mathcal{S}} \right) \right]
		\left( e^{-i \frac{\xi}{\mathcal{T}}  x' } \right) \dd{p} \dd{x'} \\
		&= \frac{1}{\sqrt{2\pi \mathcal{C}}} 
		\int \ket{p}_{P_1} \otimes \ket{ \mathcal{T} \left(p +\frac{\eta}{\mathcal{S}} \right)}_{X_2}
		\left( e^{-i \xi  \left(p +\frac{\eta}{\mathcal{S}} \right) } \right) \dd p.
	\end{split}
\end{equation}
Thus,
\begin{equation}
	\hat{B}' \ket{\xi, \eta} 
    = [- \mathcal{S}(\phi) \hat{P}_1 +  \mathcal{C}(\phi) \hat{X}_2] \ket{\xi, \eta}  
    = - \mathcal{S} p + \mathcal{C} \mathcal{T} \left(p +\frac{\eta}{\mathcal{S}} \right) \ket{\xi, \eta}
	=\eta \ket{\xi, \eta}, 
\end{equation} 
via relation $\mathcal{S} = \mathcal{T} \mathcal{C}$.
Besides, the normalization of state $\ket{\xi, \eta}$ is shown as
\begin{equation}
	\begin{split}
		\langle \xi', \eta' | \xi, \eta \rangle
		&=\frac{1}{2\pi |\mathcal{T} \mathcal{S}|} 
		\iint \dd{x} \dd{x'} e^{\frac{i}{\mathcal{S} } (\eta' x'- \eta x) } 
		\delta (x-x') \delta \qty[\frac{(x-\xi)-(x'-\xi')}{\mathcal{T}}] \\
		&=\frac{1}{2\pi |\mathcal{T} \mathcal{S}|} 
		\int \dd{x} e^{\frac{i}{|\mathcal{S}| } (\eta'- \eta) x } 
		\delta \qty(\frac{\xi'-\xi}{\mathcal{T}})
		=\delta (\xi'-\xi) \frac{1}{2\pi |\mathcal{S}|}  
		\int \dd{x} e^{\frac{i}{\mathcal{S} } (\eta'- \eta) x }  \\
		&= \delta (\xi'-\xi) \delta (\eta'-\eta).
	\end{split}
\end{equation}

\subsection{the joint probability density function}\label{appd:PDF}
According to the coherent state $\ket{\alpha}$, with real and imaginary parts $\vartheta_1$ and $\vartheta_2$, in the coordinate and momentum representations,
\begin{equation}
	\begin{split}
		\psi_\alpha(x)
		&= _X\braket{x}{\alpha}
		= \frac{1}{\pi^{1/4}} \exp \left[
		-\frac{(x-\sqrt{2} \vartheta_1)^2}{2} + i \sqrt{2} \vartheta_2 x
		\right], \\
		\varphi_\alpha(p)
		&=_{P}\braket{p}{\alpha}
		=\frac{1}{\pi^{1/4}} \exp \left[
		-\frac{(p-\sqrt{2} \vartheta_2)^2}{2} - i \sqrt{2} \vartheta_1 p
		\right],
	\end{split}
\end{equation}
therefore the overlap between simultaneous eigenstate \eqref{eq:simultaneous eigenstate} and the two-mode coherent state $\ket{\alpha_1, \alpha_2}$ is given by
\begin{equation}
	\begin{split}
        \braket{\xi,\eta}{\alpha_1, \alpha_2}
		= \frac{1}{\sqrt{2\pi \mathcal{T} \mathcal{S}}}
		\int \dd{x} e^{i\frac{\eta}{\mathcal{S}} x} 
		\psi_{\alpha_1} (x) 
		\varphi_{\alpha_2} \left(\frac{x-\xi}{\mathcal{T}}\right).
	\end{split}
\end{equation}
Applying the coherent state amplitudes $\alpha_1$ and $\alpha_2$ in Eqs.~\eqref{eq:alphas} (or their real and imaginary parts in Eq.~\eqref{eq:ReIm of alpha} directly) to the above equation, we can obtain
\begin{equation}\label{eq:inner product}
\begin{split}
    \braket{\xi,\eta}{\alpha_1, \alpha_2}
    =&\frac{1}{\sqrt{2\pi \mathcal{T} \mathcal{S}}} \times \\
    &\int \dd{x} e^{i\frac{\eta}{\mathcal{S}} x} 
    \left\{ \frac{1}{\pi^{1/4}} \exp\left[ -\frac{(x-A')^2}{2} + i B' \mu x \right]     
    \cdot
    \frac{1}{\pi^{1/4}} \exp\left[ -\frac{1}{2} \left( \frac{x-\xi}{\mathcal{T}} \right)^2 - iB'\sqrt{1-\mu^2} \left( \frac{x-\xi}{\mathcal{T}} \right)  \right] \right\}.
\end{split}
\end{equation}
This integral of a Gaussian function can be easily obtained and, subsequently, the joint probability density function is given by
\begin{equation}
	p(\xi, \eta) 
	=|\langle  \xi,\eta | \alpha_1, \alpha_2 \rangle|^2
	=\frac{1}{\pi(1+\mathcal{T}^2)} \left| \frac{\mathcal{T}}{\mathcal{S}} \right| 
	\exp\left\{
	-\frac{\mathcal{S}^2 (\xi-A')^2 + \left[\mathcal{T} \eta - \mathcal{S} \left( \sqrt{1-\mu^2} -\mathcal{T} \mu  \right) B ' \right]^2 }{\mathcal{S}^2(1+\mathcal{T}^2)}
	\right\}.
\end{equation}
This result as a Gaussian function of two parameters $\xi$ and $\eta$ can be inserted into Eq.~\eqref{eq:CFIM} to obtain the classical FIM with respect to the scaled parameters $A' = \mathcal{N} A$ and $B' = \mathcal{N} B$ as
\begin{equation}\label{eq:mat cF for scaled para}
	2 \left[
	\begin{array}{cc}
	\cos^2\phi   	& 0 \\
	0   			& \left(\sqrt{1-\mu^2} \cos \phi - \mu \sin \phi\right)^2
	\end{array}
	\right].
\end{equation}
Thus, as Eq.~\eqref{eq:mat cF} in the main text, the classical FIM
with respect to the parameters $A$ and $B$ should be
\begin{equation}
	F = 2 \mathcal{N}^2 \left[
	\begin{array}{cc}
	\cos^2\phi   	& 0 \\
	0   			& \left(\sqrt{1-\mu^2} \cos \phi - \mu \sin \phi\right)^2
	\end{array}
	\right].
\end{equation}

\subsection{the saturation of the measurement protocol}\label{appd:condition}
Here we derive the saturation of the measurement Eq.~\eqref{eq:measurement} and the corresponding condition Eq.~\eqref{eq:tight condition}. 
Substituting classical FIM Eq.~\eqref{eq:mat cF} and matrix $\mathcal{Q}$ Eq.~\eqref{eq:mat Q} into the definition of normalized-square-root regret of Fisher information, we obtain
\begin{equation}
	\Delta_1 = \sqrt{\frac{\Re \mathcal{Q}_{11} - F_{11}}{\Re \mathcal{Q}_{11}}}
	=\sqrt{\frac{2-2\cos^2\phi}{2}}
	=\left|\sin \phi \right|,
\end{equation}
and
\begin{equation}
	\begin{split}
		\Delta_2 &= \sqrt{\frac{\Re \mathcal{Q}_{22} - F_{22}}{\Re \mathcal{Q}_{22}}}
		= \sqrt{\frac{2 - 2(\sqrt{1-\mu^2}\cos \phi - \mu \sin \phi)^2 }{2}} \\
		&= \left|\mu \cos \phi + \sqrt{1-\mu^2} \sin \phi \right|. 
	\end{split}
\end{equation}
Based on above normalized-square-root regrets, the left side of the tradeoff relation \eqref{eq:IRTR} becomes
\begin{equation}\label{eq:left hand side}
	\begin{split}
		L = \sin^2 \phi + \left( \mu \cos \phi + \sqrt{1-\mu^2} \sin \phi \right)^2  
		+ 2\sqrt{1-\mu^2} \sin \phi \left|\mu \cos \phi + \sqrt{1-\mu^2} \sin \phi \right|,
	\end{split}
\end{equation}
with $0 \leq \phi \leq \pi$ and $c_{12} = \mu$ based on Eq.~ \eqref{eq:incompat coeff}.

If the condition which we establish in Eq.~\eqref{eq:tight condition} is satisfied, i.e.,
$$\mu \cos \phi + \sqrt{1-\mu^2} \sin \phi \leq 0, $$ 
the above expression \eqref{eq:left hand side} can be simplified as
\begin{equation}
	L = \mu^2. 
\end{equation}
On the contrary, if $\mu \cos \phi + \sqrt{1-\mu^2} \sin \phi > 0$, the left side of the tradeoff relation \eqref{eq:left hand side} turns to
\begin{equation}
	L = \mu^2 + 4 \sqrt{1-\mu^2} \sin \phi
	\left(\sqrt{1-\mu^2} \sin \phi + \mu \cos \phi \right). 
\end{equation}
To sum up, the left side of the tradeoff relation \eqref{eq:IRTR} or \eqref{eq:CFIM tradeoff relation} is given by
\begin{equation}
	L=
	\begin{cases}
		\mu^2 & \mu \cos \phi + \sqrt{1-\mu^2} \sin \phi \leq 0, \\
		\mu^2 + 4 \sqrt{1-\mu^2} \sin \phi
		\left(\sqrt{1-\mu^2} \sin \phi + \mu \cos \phi \right) & \mu \cos \phi + \sqrt{1-\mu^2} \sin \phi > 0.
	\end{cases}
\end{equation}
Under the specific condition Eq.~\eqref{eq:tight condition}, the measurement Eq.~\eqref{eq:measurement} is proved to saturate the ultimate tradeoff relation \eqref{eq:IRTR} or \eqref{eq:CFIM tradeoff relation}.

\begin{figure}[tbhp]
\includegraphics[width=0.5\columnwidth]{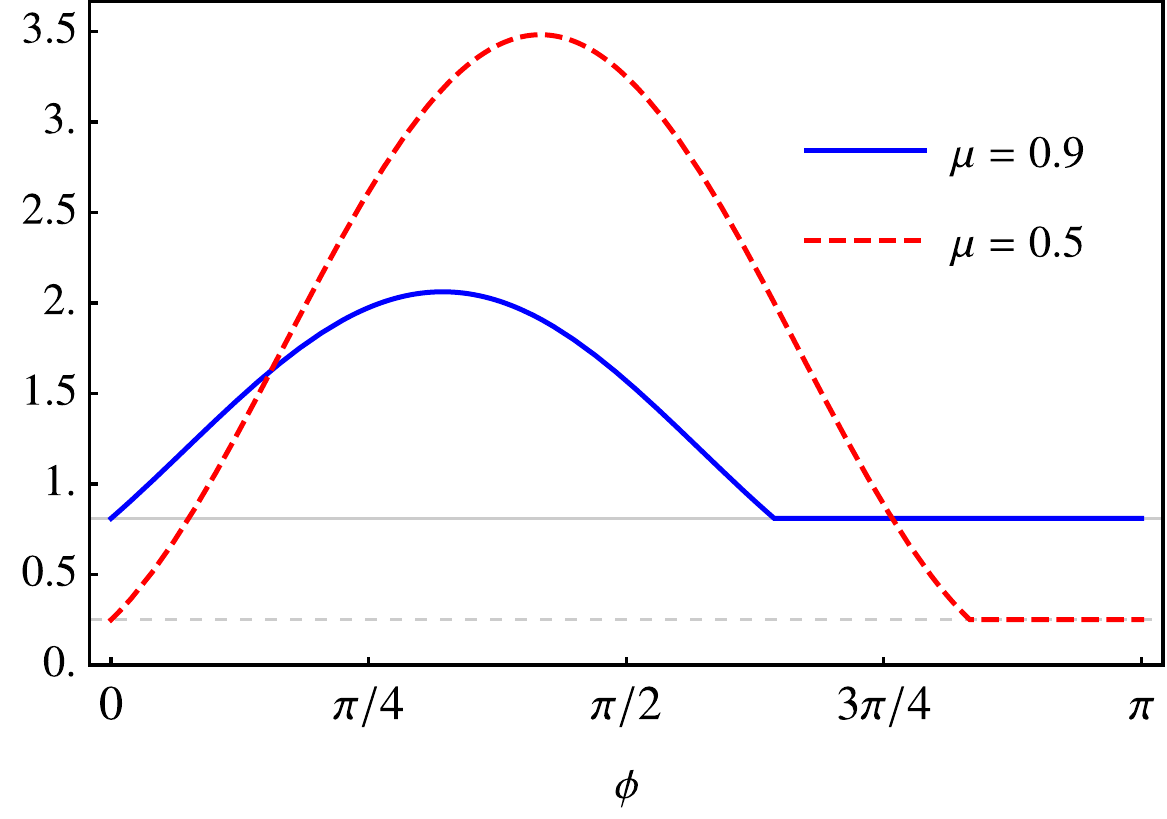}
	\centering
	\caption{
		Combined with classical FIM \eqref{eq:CFIM}, the left side of tradeoff relation \eqref{eq:CFIM tradeoff relation} as a function of $\phi$ for $\mu = 0.9$ (blue solid line) and $\mu=0.5$ (red dashed line).
		For comparison, the corresponding straight lines denote $\mu^2$, the right side of Eq.~\eqref{eq:CFIM tradeoff relation}.
		The flat parts mean that tradeoff relation \eqref{eq:CFIM tradeoff relation} becomes equality and tight.
	}
	\label{fig:Phi Cond}
\end{figure}

We utilize the classical FIM \eqref{eq:mat cF} to compute and plot the left side of tradeoff relation \eqref{eq:CFIM tradeoff relation} versus $\phi$ in Fig.~\ref{fig:Phi Cond}.
The flat parts of both curves for two various $\mu$ indicate that the inequality relation \eqref{eq:CFIM tradeoff relation} becomes an equality when $\phi$ satisfies condition \eqref{eq:tight condition}.
Remarkably, our tradeoff relation \eqref{eq:CFIM tradeoff relation} is valuable to certify that the joint measurement of the observables~\eqref{eq:measurement} saturates the inequality \eqref{eq:CFIM tradeoff relation}, and fully provides accessible region of estimation precisions.

\section{Estimating Signals with Squeezing State}\label{appd:sqz}
\subsection{quantum FIM}
Here we consider the scenario of using a squeezed vacuum state to estimate the signal $s(t) = A \cos \Omega t + B \sin \Omega t$.
If the initial state is squeezed through the squeeze operator $S (r) = \exp[\frac{r}{2} (a^2 - a^{\dagger2})]$, with annihilation (creation) operator $a$ ($a^\dagger$) of any mode and squeezing parameter $r \in \mathbb{R}$, the quadrature vector Eq.~\eqref{eq:vec x with unit vector} or \eqref{eq:expand vec x with unit vector} is transformed as
\begin{align}\label{eq:sqz vec x}
	 {\mathbf x}
    = S^\dagger (r) \mathbf{x}^{(0)} S(r)
    + A \mathbf{d}_A + B \mathbf{d}_B
    = \mathcal{R} (-r) {\mathbf x}^{(0)} 
    + A \mathbf{d}_A + B \mathbf{d}_B
    = \mqty [
    x_1^{(0)} e^{-r} \\
    p_1^{(0)} e^{r} \\
    x_2^{(0)} e^{-r}\\
    p_2^{(0)} e^{r}
    ] 
    + A \mathbf{d}_A + B \mathbf{d}_B.
\end{align}
with the relation
$   S^\dagger (r) \mathbf{x}^{(0)} S(r) = \mathcal{R} (-r)  \mathbf{x}^{(0)}$
and diagonal matrix 
$\mathcal{R} (r) = \mathrm{diag}(e^{r}, e^{-r}, e^{r}, e^{-r})$,
where the quadrature vector 
${\mathbf x}^{(0)} = \qty(x_1^{(0)}, p_1^{(0)}, x_2^{(0)}, p_2^{(0)})  $ still
represents initial state without squeezing.
If the initial state is a vacuum state $\ket{0} \otimes \ket{0}$, the covariance matrix of the new quadratures in the vector $\mathbf{x} = \qty(x_1, p_1, x_2, p_2)$ is given by
$ \mathrm{diag} \qty(\frac{e^{-2r}}{2}, \frac{e^{2r}}{2}, \frac{e^{-2r}}{2}, \frac{e^{2r}}{2})$,
and the corresponding state to be measured is a two-mode displaced squeezed state
\begin{equation}
\ket{\beta_1, \beta_2 ; r} 
:= D(\beta_1) S(r)\ket{0} \otimes  D(\beta_2) S(r)\ket{0} 
= S(r) D(\widetilde{\beta}_1) \ket{0} \otimes S(r) D(\widetilde{\beta}_2) \ket{0}
\end{equation}
with
\begin{equation}\label{eq:sqz beta}
 \widetilde{\beta}_j = \beta_j \cosh r + \beta_j^\ast \sinh r.
\end{equation}
Since $S(r)$ is a unitary operator independent of the estimated parameters, the quantum geometric tensor for the parametric state $\ket{\beta_1, \beta_2 ; r}$ equals that for the two-mode coherent state 
$ \ket{\widetilde{\beta}_1, \widetilde{\beta}_2} = D(\widetilde{\beta}_1) \ket{0} \otimes D(\widetilde{\beta}_2) \ket{0}$.
According to Eq.~\eqref{eq:sqz beta}, the real and imaginary parts of 
$\widetilde{\beta}_j = \widetilde{\beta}_j^\mathrm{r} + i \widetilde{\beta}_j^\mathrm{i}$, $\widetilde{\beta}_j^\mathrm{r}$ and $\widetilde{\beta}_j^\mathrm{i}$, are related with the ones of $\beta_j$ as
\begin{align}\label{eq:trans sqz beta to normal beta}
	\mqty [
    \widetilde{\beta}_1^\mathrm{r} \\
    \widetilde{\beta}_1^\mathrm{i} \\
    \widetilde{\beta}_2^\mathrm{r} \\
    \widetilde{\beta}_2^\mathrm{i}
    ] 
    = \mqty[\dmat{e^{r}, e^{-r}, e^{r}, e^{-r}}]
    \mqty [
    {\beta}_1^\mathrm{r} \\
    {\beta}_1^\mathrm{i} \\
    {\beta}_2^\mathrm{r} \\
    {\beta}_2^\mathrm{i}
    ]
    = \mathcal{R} (r)
    \mqty [
    {\beta}_1^\mathrm{r} \\
    {\beta}_1^\mathrm{i} \\
    {\beta}_2^\mathrm{r} \\
    {\beta}_2^\mathrm{i}
    ].
\end{align}
Based on the relation \eqref{eq:trans para to beta}, it means
\begin{align}\label{eq:ReIm of sqz beta}
	\mqty [
    \widetilde{\beta}_1^\mathrm{r} \\
    \widetilde{\beta}_1^\mathrm{i} \\
    \widetilde{\beta}_2^\mathrm{r} \\
    \widetilde{\beta}_2^\mathrm{i}
    ] 
    &= \mathcal{R} (r) \mathcal{P} \mqty [A \\ B]
    = \frac{\sqrt{T}}{2} \mqty[\dmat{e^{r}, e^{-r}, e^{r}, e^{-r}}]
	\mqty[
	\Re[\chi_{xG}] & - \Im[\chi_{xG}] \\
	\Re[\chi_{pG}] & - \Im[\chi_{pG}] \\
	\Im[\chi_{xG}] &   \Re[\chi_{xG}] \\
	\Im[\chi_{pG}] &   \Re[\chi_{pG}]
	]
    \mqty [A \\ B] \notag \\
    &= \frac{\sqrt{T}}{2} 
	\mqty[
	e^{r} \Re[\chi_{xG}] & - e^{r} \Im[\chi_{xG}] \\
	e^{-r} \Re[\chi_{pG}] & - e^{-r}\Im[\chi_{pG}] \\
	e^{r} \Im[\chi_{xG}] &   e^{r} \Re[\chi_{xG}] \\
	e^{-r} \Im[\chi_{pG}] &   e^{-r} \Re[\chi_{pG}]
	]
    \mqty [A \\ B]
    = \frac{\sqrt{T}}{2} \mqty[
    e^{r} (\Re[\chi_{xG}] A - \Im[\chi_{xG}] B)   \\
    e^{-r} (\Re[\chi_{pG}] A -\Im[\chi_{pG}] B) \\
    e^{r} (\Im[\chi_{xG}] A + \Re[\chi_{xG}] B)  \\
    e^{-r} (\Im[\chi_{pG}] A +  \Re[\chi_{pG}] B)
    ].
\end{align}
Since the quantum geometric tensor with respect to the real and imaginary parts
$(\widetilde{\beta}_1^\mathrm{r}, \widetilde{\beta}_1^\mathrm{i}, \widetilde{\beta}_2^\mathrm{r}, \widetilde{\beta}_2^\mathrm{i})$ for the two-mode coherent state $\ket{\widetilde{\beta}_1, \widetilde{\beta}_2}$ is still $\mathcal{Q}_0$ in Eq.~\eqref{eq:Q for ReIm part}, 
the one with respect to the estimated parameters $A$ and $B$ is given by
\begin{align}
    \mathcal{Q}_r = \mathcal{P}^\mathrm{T} \mathcal{R}^\mathrm{T} (r) \mathcal{Q}_0 \mathcal{R} (r) \mathcal{P} &= T
    \mqty[
	e^{2r} |\chi_{xG}|^2 + e^{-2r} |\chi_{pG}|^2 & 2i (\Re \chi_{pG} \Im \chi_{xG} - \Re \chi_{xG} \Im \chi_{pG}) \\
   -2 i (\Re \chi_{pG} \Im \chi_{xG} - \Re \chi_{xG} \Im \chi_{pG}) & e^{2r} |\chi_{xG}|^2 + e^{-2r} |\chi_{pG}|^2  
	] \notag \\
    &= 2 \mathcal{N}_r^2 
    \mqty(
	1        & i \mu_r \\
   - i \mu_r & 1  
	),
\end{align}
with the modified normalization factor $\mathcal{N}_r$ and the incompatibility coefficient $\mu_r$ as
\begin{align}
\mathcal{N}_r &= \sqrt{ \frac{T}{2} \qty(e^{2r} |\chi_{xG}|^2 + e^{-2r} |\chi_{pG}|^2) }, \notag \\
\mu_r &= \qty(\frac{\mathcal{N}}{\mathcal{N}_r})^2 \mu
= \frac{\Re \chi_{pG} \Im \chi_{xG} - \Re \chi_{xG} \Im \chi_{pG}}{\mathcal{N}_r^2}.
\end{align}
That is, the precision tradeoff relation with squeezing should be obtained through modifying Eq.~\eqref{eq:error tradeoff relation} with replacement Eq.~\eqref{eq:replace} as
\begin{equation}\label{eq:sqz error tradeoff relation}
\qty[2 - \frac{ (\mathcal{N}_r^2 \mathcal{E}_{A}) ^{-1}+ (\mathcal{N}_r^2 \mathcal{E}_{B}) ^{-1}}{2}] 
+ 2 \sqrt{1-\mu_r^2} \sqrt{
	\qty[1 - \frac{ (\mathcal{N}_r^2 \mathcal{E}_{A})^{-1}}{2} ]
	\qty[1 - \frac{ (\mathcal{N}_r^2 \mathcal{E}_{B})^{-1}}{2} ]
	}
\geq \mu_r^2.
\end{equation}

\subsection{classical FIM}
On the other hand, an optimal measurement that makes the above tradeoff relation tight can be obtained.
If the initial state in \eqref{eq:sqz vec x} is a vacuum state $\ket{0} \otimes \ket{0}$, the quadrature vector $\mathbf{x} = (x_1, p_1, x_2, p_2)^\mathrm{T}$ in \eqref{eq:sqz vec x} is equivalent to the other operator vector $\hat{\mathbf{x}} = (\hat{x}_1, \hat{p}_1, \hat{x}_2, \hat{p}_2)^\mathrm{T}$ for acting on a two-mode squeezed state
$\ket{\beta_1, \beta_2 ; r} 
= S(r) D(\widetilde{\beta}_1) \ket{0} \otimes S(r) D(\widetilde{\beta}_2) \ket{0} = S(r) \otimes S(r) \ket{\widetilde{\beta}_1, \widetilde{\beta}_2}$, with a two-mode coherent state $\ket{\widetilde{\beta}_1, \widetilde{\beta}_2}$, in the Schr\"{o}dinger picture. 
Moreover, this linear framework is equivalent to the scenario where a modified quadrature vector 
\begin{equation}\label{eq:sqz vec x in Schr}
    \hat{\widetilde{\mathbf x}} = S^\dagger (r) \hat{\mathbf x} S(r)
    = \mathcal{R} (-r) \hat{\mathbf x}
    = (e^{-r} \hat{x}_1, e^{r} \hat{p}_1, e^{-r} \hat{x}_2, e^{r} \hat{p}_2)^\mathrm{T}
\end{equation}
acts on the two-mode coherent state $\ket{\widetilde{\beta}_1, \widetilde{\beta}_2}$.
Based on Eq.~\eqref{eq:sqz vec x}, the modified vector $\hat{\widetilde{\mathbf x}}$  in the Schr\"{o}dinger picture can be shifted to the quadrature vector in the Heisenberg picture, 
\begin{align}\label{eq:modified vec x}
\widetilde{\mathbf{x}} 
= \Omega^\mathrm{T} \mathcal{R} (-r) \Omega
= \mathcal{R} (r) \mathbf{x}
= \mathbf{x}^{(0)} 
+ A \widetilde{\mathbf{d}}_A 
+ B \widetilde{\mathbf{d}}_B,
\end{align}
for acting on the vacuum state $\ket{0,0}$, where
$\widetilde{\mathbf{d}}_A:=
\mathcal{R} (r) \mathbf{d}_A =
\sqrt{\frac{T}{2}} \qty(
e^{r} \Re [\chi_{xG}], 
e^{-r} \Re [\chi_{pG}], 
e^{r} \Im [\chi_{xG}], 
e^{-r} \Im [\chi_{pG}] )^\mathrm{T}$
and
$\widetilde{\mathbf{d}}_B:= 
\mathcal{R} (r) \mathbf{d}_B =
\sqrt{\frac{T}{2}} \qty(
- e^{r} \Im [\chi_{xG}], 
- e^{-r} \Im [\chi_{pG}], 
e^{r} \Re [\chi_{xG}], 
e^{-r} \Re [\chi_{pG}] )^\mathrm{T}$
are two modified orthogonal vectors with the common Euclidean norm $\mathcal{N}_r$, and the $4 \times 4$ matrix $\Omega$ is defined in Eq.~\eqref{eq:Omega transfer}.
A symplectic transformation matrix $\mathcal{M}_r$, analogous to $\mathcal{M}$ in Eq.~\eqref{eq:vec X}, can transform $\widetilde{\mathbf{x}}$, i.e., \eqref{eq:modified vec x}, into
\begin{equation}\label{eq:sqz vec X}
    \widetilde{{\mathbf X}} 
    = \left[\begin{array}{c}
         \widetilde{X}_1  \\
         \widetilde{P}_1  \\
         \widetilde{X}_2  \\
        \widetilde{P}_2
    \end{array}\right]
    = \mathcal{M}_r \widetilde{\mathbf{x}}
    = \widetilde{\mathbf X}^{(0)} + A \mathcal{N}_r
    \left[\begin{array}{c}
         1  \\
         0  \\
         0  \\
         0
    \end{array}\right]  + B \mathcal{N}_r
    \left[\begin{array}{c}
         0                 \\
         \mu_r             \\
         \sqrt{1-\mu_r^2}  \\
         0
    \end{array}\right] ,
\end{equation}
in the same form as \eqref{eq:vec X}.
Therefore, the vector $\widetilde{\mathbf{X}}$ in Eq.~\eqref{eq:sqz vec X} corresponds to a two-mode coherent state $\ket{\widetilde{\alpha}_1, \widetilde{\alpha}_2}$ with 
$\widetilde{\alpha}_1 = \frac{\mathcal{N}_r}{\sqrt{2}} \left(A + i \mu_r B \right)$
and
$\widetilde{\alpha}_2 = \frac{\mathcal{N}_r}{\sqrt{2}} \sqrt{1-\mu_r^2} B$.
Then the optimal compatible unbiased estimates for $A$ and $B$ is similar to Eq.~\eqref{eq:measurement} or \eqref{eq:measurement in SM} as
\begin{align}\label{eq:sqz measurement in SM}
	\hat{A} = \mathcal{N}_r^{-1} 
    [\widetilde{X}_1 - \mathcal{T}(\phi) \widetilde{P}_2], \quad 
	\hat{B} = \mathcal{N}_r^{-1}
    [- \mathcal{S}_r (\phi) \widetilde{P}_1 +  \mathcal{C}_r (\phi) \widetilde{X}_2],
\end{align}
with modified coefficients $ \mathcal{C}_r (\phi) = \cos \phi / \left( \sqrt{1-\mu_r^2} \cos \phi - \mu_r \sin \phi \right)$ and $\mathcal{S}_r = \sin \phi / \left( \sqrt{1-\mu_r^2} \cos \phi - \mu_r \sin \phi \right)$.
Eventually, owing to the similar form, the corresponding classical FIM can be obtained by replacing the coefficients $\mathcal{N} \rightarrow \mathcal{N}_r$ and $\mu \rightarrow \mu_r$ in Eq.~\eqref{eq:mat cF} as
\begin{equation}
	F_r = 2 \mathcal{N}_r^2 \left[
	\begin{array}{cc}
		\cos^2\phi   	& 0 \\
		0   					& \left(\sqrt{1-\mu_r^2} \cos \phi - \mu_r \sin \phi\right)^2
	\end{array}
	\right].
\end{equation}
Moreover, the condition for saturating the tradeoff relation Eq.~\eqref{eq:sqz error tradeoff relation} is given by
\begin{equation}
	\mu_r \cos \phi + \sqrt{1-\mu_r^2} \sin \phi \leq 0, 
\end{equation}
via the replacement $\mu \rightarrow \mu_r$ in Eq.~\eqref{eq:tight condition}.

Based on the linear relation 
${\widetilde{\mathbf X}} = \mathcal{M}_r \widetilde{\mathbf x}$
in \eqref{eq:sqz vec X} and Eq.~\eqref{eq:Omega transfer},
the operator vector $\hat{\widetilde{\mathbf X}} = (\hat{\widetilde{X}}_1, \hat{\widetilde{P}}_1, \hat{\widetilde{X}}_2, \hat{\widetilde{P}}_2)$ in the Schr\"{o}dinger picture that act on the two-mode coherent state 
$\ket{\widetilde{\alpha}_1, \widetilde{\alpha}_2}$ 
should be
\begin{equation}
    \hat{\widetilde{\mathbf X}}
    = \Omega^{T} \mathcal{M}_r^\mathrm{T} \Omega \hat{\widetilde{{\mathbf x}}}
    = \Omega^{T} \mathcal{M}_r^\mathrm{T} \Omega (e^{-r} \hat{x}_1, e^{r} \hat{p}_1, e^{-r} \hat{x}_2, e^{r} \hat{p}_2 )^\mathrm{T},
\end{equation}
combined with Eq.~\eqref{eq:sqz vec x in Schr}.
Back to the Schr\"{o}dinger picture, the optimal compatible unbiased estimates in \eqref{eq:sqz measurement in SM} can be acquired straightforwardly by replacing the operators $\widetilde{X}_j$ and $\widetilde{P}_j$ with another ones $\hat{\widetilde{X}_j}$ and $\hat{\widetilde{P}_j}$ in the vector $\hat{\widetilde{\mathbf X}}$.

\end{document}